\documentstyle[12pt,aaspp4]{article}

\begin{document}

%
\def\arcsec{\ifmmode '' \else $''$\fi}
\def\arcmin{$'$}
\def\arcsecpoint{\ifmmode ''\!. \else $''\!.$\fi}
\def\arcminpoint{$'\!.$}
\def\ltsim{\raisebox{-.5ex}{$\;\stackrel{<}{\sim}\;$}}
\def\gtsim{\raisebox{-.5ex}{$\;\stackrel{>}{\sim}\;$}}
\def\hi{H\,{\sc i}}
\def\hii{H\,{\sc ii}}
\def\hei{He\,{\sc i}}
\def\heii{He\,{\sc ii}}
\def\cii{C\,{\sc ii}}
\def\ciii{C\,{\sc iii}}
\def\civ{C\,{\sc iv}}
\def\ni{N\,{\sc i}}
\def\nii{N\,{\sc ii}}
\def\nv{N\,{\sc v}}
\def\oi{O\,{\sc i}}
\def\oii{O\,{\sc ii}}
\def\oiii{O\,{\sc iii}}
\def\oiv{O\,{\sc iv}}
\def\ovi{O\,{\sc vi}}
\def\neiii{Ne\,{\sc iii}}
\def\neiv{Ne\,{\sc iv}}
\def\nev{Ne\,{\sc v}}
\def\nevi{Ne\,{\sc vi}}
\def\navi{Na\,{\sc vi}}
\def\mgii{Mg\,{\sc ii}}
\def\mgv{Mg\,{\sc v}}
\def\mgvii{Mg\,{\sc vii}}
\def\mgviii{Mg\,{\sc viii}}
\def\alv{Al\,{\sc v}}
\def\alvi{Al\,{\sc vi}}
\def\alviii{Al\,{\sc viii}}
\def\sii{Si\,{\sc ii}}
\def\siii{Si\,{\sc iii}}
\def\siiv{Si\,{\sc iv}}
\def\sivi{Si\,{\sc vi}}
\def\sivii{Si\,{\sc vii}}
\def\siix{Si\,{\sc ix}}
\def\six{Si\,{\sc x}}
\def\suii{S\,{\sc ii}}
\def\suiii{S\,{\sc iii}}
\def\suviii{S\,{\sc viii}}
\def\suix{S\,{\sc ix}}
\def\suxi{S\,{\sc xi}}
\def\ariii{Ar\,{\sc iii}}
\def\arvi{Ar\,{\sc vi}}
\def\arx{Ar\,{\sc x}}
\def\arxi{Ar\,{\sc xi}}
\def\caii{Ca\,{\sc ii}}
\def\caviii{Ca\,{\sc viii}}
\def\feii{Fe\,{\sc ii}}
\def\fevii{Fe\,{\sc vii}}
\def\fex{Fe\,{\sc x}}
\def\fexi{Fe\,{\sc xi}}
\def\fexiv{Fe\,{\sc xiv}}
\def\n{\footnotemark}
\def\o{\o}
%

\title{Locally Optimally-emitting Clouds and the Narrow
Emission Lines in Seyfert Galaxies}

\author{Jason W. Ferguson, Kirk T. Korista}
\affil{Department of Physics \& Astronomy, University of Kentucky,
Lexington, KY 40506}

\author{Jack A. Baldwin}
\affil{Cerro Tololo Interamerican Observatory\altaffilmark{1}, Casilla
603, La Serena, Chile}

\author{Gary J. Ferland}
\affil{Department of Physics \& Astronomy, University of Kentucky,
Lexington, KY 40506}

\altaffiltext{1}{Operated by the Association of Universities for
Research in Astronomy Inc.\ (AURA) under cooperative agreement with the
National Science Foundation}

\begin{abstract}
The narrow emission line spectra of active galactic nuclei are not
accurately described by simple photoionization models of single
clouds.  Recent Hubble Space Telescope images of Seyfert 2 galaxies
show that these objects are rich with ionization cones, knots,
filaments, and strands of ionized gas.  Here we extend to the narrow
line region the ``locally optimally emitting cloud'' (LOC) model, in
which the observed spectra are predominantly determined by powerful
selection effects.  We present a large grid of photoionization models
covering a wide range of physical conditions and show the optimal
conditions for producing many of the strongest emission lines.  We show
that the integrated narrow line spectrum can be predicted by an
integration of an ensemble of clouds, and we present these results in
the form of diagnostic line ratio diagrams making comparisons with
observations. We also predict key diagnostic line ratios as a function
of distance from the ionizing source, and compare these to
observations. The predicted radial dependence of the [\oiii\/]/[\oii\/]
ratio may be matched to the observed one in NGC~4151, if the narrow
line clouds see a more intense continuum than we see. The LOC scenario
when coupled with a simple Keplerian gravitational velocity field will
quite naturally predict the observed line width versus critical density
relationship.  The influence of dust within the ionized portion of the
clouds is discussed and we show that the more neutral gas is likely to
be dusty, although a high ionization dust-free region is most likely
present too.  This argues for a variety of NLR cloud origins.
\end{abstract}


\newpage

\section{Introduction}
The failure of ``simple'' photoionization models to describe the
narrow emission line spectrum of active galactic nuclei (AGN) was
shown dramatically by Filippenko \& Halpern (1984), who found
that the line widths of the optical forbidden lines were correlated
with the critical density and ionization potential of those lines.
This correlation indicates stratification of the photoionized
region, with the lines emitting near their critical densities.

Recent Hubble Space Telescope images of Seyfert 2 galaxies show that
these objects are rich with ionization cones (Evans et al.\ 1991;
Tsvetanov \& Walsh 1992; Wilson et al.\ 1993; Macchetto et al.\ 1994;
and Arribas et al.\ 1996), knots (Acosta-Pulid\'{o} et~al. 1996), and
filaments, and strands (Falcke et~al.  1996) of ionized gas.  We should
not expect that these complex environments can be described with
simple, single-component photoionization models.  Similar images of
Seyfert 1 type objects show a less complex, symmetric structure that
may indicate a viewing angle preference consistent with the unification
model of Seyfert galaxies (Schmitt \& Kinney 1996).

Two-component photoionization models have been proposed by Binette,
Wilson, \& Storchi-Bergmann (1996) and have met with some success in
describing the emission line spectrum of Seyfert galaxies.  This model
assumes that the narrow line region (NLR) is composed of both
matter-bounded and ionization-bounded clouds, and that the covering
fraction of the two populations can account for the range of observed
line emission.  A multi-component photoionization model has been
proposed by Komossa \& Schulz (1997), who assume that the NLR is
effectively composed of a few clouds with a range of gas densities and
distances from the central AGN.  Both of these types of models do much
better at reproducing the scatter in the line ratio observations than
single component photoionization models.

An alternative to photoionization by the central source is
photoionization by fast moving shocks.  Dopita \& Sutherland (1995)
describe in detail these ``photoionizing'' shocks and argue that all
narrow emission lines can be excited by these processes.  However,
Morse, Raymond, \& Wilson (1996) and Wilson (1997) discuss the
viability of such shocks in describing the emission spectrum of active
galaxies.  In particular, Wilson (1997) argues that the energetics of
radiative shocks may not account for the observed line luminosities.
It is still unclear exactly what role shocks play in the NLR
environment; that is, whether they provide the bulk of the energy input
for the observed emission or merely compress gas so that it emits more
strongly when photoionized by a central source (Pogge 1997).

We will assume that photoionization by the central source is the
dominating mechanism responsible for NLR emission.  We extend the
locally optimally-emitting clouds (LOC) hypothesis of Baldwin et~al.
(1995) to NLR gas.  This assumes that in nature, gas with a wide range
of physical conditions is present and the emission line spectrum we
observe is predominately a result of a simple, yet powerful, selection
effect:  we observe lines from clouds best able to emit them.  To this
end we present the results of a large grid of photoionization models
and make predictions of the nature of the integrated spectrum.

\section{Model calculations}

\subsection{Assumptions}
For simplicity we will assume that the source of narrow line emission
observed in active galaxies is from photoionized clouds that are
similar to illuminated molecular clouds in our own galaxy.  The source
of these clouds may be four-fold:  (1) molecular cloud complexes in the
disk of the galaxy; (2) the dusty molecular torus that is believed to
surround the central source and which will certainly be photoionized;
(3) ISM entrained behind a radio jet so that photoionization by the
central source becomes important; and (4) material ``snow plowed'' in
the form of bow shocks in front of a radio jet such that it too may be
photoionized by the central source.  We will ignore emission that may
be due to shock excitation coming from photo-ionizing shocks, although
how important this emission is to the overall spectrum is unclear
(Wilson 1997).

We have used the spectral synthesis code {\sc Cloudy} (version 90.03;
Ferland 1996) to calculate the emission from plane parallel, constant
hydrogen density clouds ionized by a continuum similar to that expected
from a typical Seyfert galaxy with $L_{ion} =
10^{43.5}$~ergs~s$^{-1}$.  The shape of the ionizing continuum was
chosen to be a combination of a UV-bump of the form $f_{\nu} \propto
\nu^{-0.3} exp(-h\nu /kT_{cut})$ and an X-ray power law of the form
$f_{\nu} \propto \nu^{-1.0}$ spanning 13.6~eV to 100~keV.  The UV-bump
cutoff temperature, $T_{cut}$, was chosen such that the UV-bump peaked
(in $\nu F _{\nu}$) at 48~eV.  The UV and X-ray components were
combined with a typical Seyfert UV to X-ray spectral slope,
$\alpha_{ox} = -1.2$.

The ionization/thermal equilibrium and radiative transfer calculations
for a single cloud proceeded until one of the following three
conditions were met. First, the electron temperature dropped below
3000~K.  Gas with lower temperatures does not contribute significantly
to the emission lines presented in this paper.  Second, we do not allow
clouds to have thicknesses in excess of 10\% of its distance from the
central continuum source.  Third, we do not allow the total hydrogen
column density to exceed 10$^{24}$ cm$^{-2}$; this mainly affects the
very high ionization clouds and prevents them from becoming Thomson
thick.  In practice the third condition had little impact on the narrow
emission lines.  The largest cloud of the grid presented below is one
with a thickness of $\sim$\/16~pc, consistent with the size of large
molecular clouds.  We will comment further on the effects of these
stopping criteria in a later section.

Finally, we assume that each cloud sees the full continuum with no
obscuration.  At large enough distances, intervening clouds or diffuse
ISM may attenuate the ionizing spectrum significantly.  We have looked
at the effects of such ISM attenuation on the incident continuum: An
ISM with gas densities as low as 0.1~cm$^{-3}$ such as found in the
local ISM of the Milky Way (Wood \& Linsky 1997) would have no effect
on the results presented below.  A higher density ISM (n(H) =
0.5~cm$^{-3}$) however, would have an effect at large distances from
the ionizing source, due to the larger column density and lower
ionization resulting in a higher gas opacity.  For the higher density
gas and the shape and luminosity of the incident continuum given above,
the attenuated continuum has an optical depth of unity at 4~Ryd at a
distance of $\log R \sim 21.75$~cm from the ionizing source, resulting
in 5 -- 20\% changes in the line luminosities.  At larger distances the
attenuation by a dense ISM would result in a decrease in the luminosity
of the emitted spectrum of up to a factor of 10.  We neglect this
complication for simplicity.

\subsection{Reprocessing efficiency} 
We have computed, on the University of Kentucky Convex Exemplar
parallel supercomputer, a grid of 1,881 of the single cloud models
described above in a plane of hydrogen gas density and distance from
the central ionizing source.  We chose a range of parameters in order
to cover all the possible physical conditions of narrow line region
gas.  The hydrogen gas density, $n(H)$, ranged from 10$^2$ -- 10$^{10}$
cm$^{-3}$.  The incident continuum flux was varied so that for an
assumed ionizing continuum luminosity $L_{ion} = 10^{43.5}$ ergs
s$^{-1}$ (monochromatic luminosity  $\log L_{\lambda}($1450~\AA) =
39.558 ergs s$^{-1}$~\AA$^{-1}$ for the continuum shape used here) the
distance from the ionizing source of radiation, $R$, varied from
10$^{15}$ -- 10$^{22}$~cm. This distance range scales with the
luminosity as $L_{ion}^{1/2}$. The conversion from incident flux to
luminosity and radius assumes that the ionizing radiation is emitted
isotropically.  However, observations of ionizing ``cones'' (Evans et
al.\ 1991; Tsvetanov \& Walsh 1992; Wilson et al.\ 1993; Macchetto et
al.\ 1994; and Arribas et al.\ 1996) and UV-photon deficits (Binette,
Fosbury, \& Parker 1993; and Morse, Raymond \& Wilson 1996) imply that
the gas may see a different ionizing continuum than we do.

Calculations were performed assuming two abundance sets: solar and a
set including dust that varied with distance from the ionizing
source.  We will describe each set of abundances in turn and compare
the two in terms of diagnostic line ratio diagrams in a later section.

\subsubsection{Solar abundances}
Figure~1 shows the results of grid calculations, for solar abundance
clouds, in the form of contour plots of logarithm equivalent width
referred to the incident continuum at 4860~\AA\/.  The solar abundances
are from Grevesse \& Anders (1989) and Grevesse \& Noels (1993) and
have the following values:

\noindent
H :1.00E+00  He:1.00E-01  Li:2.04E-09  Be:2.63E-11  B :7.59E-10  
C :3.55E-04  N :9.33E-05  O :7.41E-04  F :3.02E-08  Ne:1.17E-04
Na:2.06E-06  Mg:3.80E-05  Al:2.95E-06  Si:3.55E-05  P :3.73E-07  
S :1.62E-05  Cl:1.88E-07  Ar:3.98E-06  K :1.35E-07  Ca:2.29E-06
Sc:1.58E-09  Ti:1.10E-07  V :1.05E-08  Cr:4.84E-07  Mn:3.42E-07
Fe:3.24E-05  Co:8.32E-08  Ni:1.76E-06  Cu:1.87E-08  Zn:4.52E-08

\noindent
Furthermore, we assume that there are no dust present in the
solar abundance computations.

The ionization parameter provides a useful homology relationship with
the emission lines (Davidson 1977).  $U(H)$ is defined as the ratio of
hydrogen ionizing photon density to hydrogen density, $U(H) \equiv
\Phi(H)/n(H)c$, where $\Phi(H)$ is the flux of hydrogen ionizing
photons and $c$ is the speed of light.  For reference a dashed line is
placed on the upper left hand plot of the first panel representing
$\log U(H) = 3.0$.  The ionization parameter increases from top right
to bottom left in these diagrams; lines of constant $U(H)$ have a slope
of $\rm{d \log}(R)/ \rm{d \log}(n) = -0.5$.

Shown in Figure~1 are 23 strong optical, UV and infrared recombination,
resonance, and forbidden lines.  The temperature of the illuminated
face of the cloud is also shown in the upper right hand plot of the
first panel.  For many emission lines a ridge of near maximum
equivalent width runs diagonally across the distance -- gas density
plane, roughly parallel to lines of constant ionization parameter.
Most lines are emitted optimally for a narrow range of ionization
parameter spanning this ridge.  For larger (smaller) ionization
parameters the gas is over (under) ionized, and the line is not
efficiently emitted. This explains the sudden drops in the emission
line equivalent widths on either side of their ridges.  Moving {\em
along} the ridge to increasing gas densities, at near constant
ionization parameter, the forbidden lines become collisionally
deexcited and the line equivalent width falls off.  Moving along the
ridge at constant ionization parameter to smaller gas densities, the
equivalent widths of some lines also diminish as other lines of similar
ionization, but lower critical density, become important coolants.
This effect is especially dramatic for \civ\/ $\lambda$1549 and \ovi\/
$\lambda$1035.

The stopping criteria assumed in $\S$\/2.1 do affect the resulting
spectrum. The grid of models shown in Figure~1 includes clouds that are
both optically thin and optically thick to hydrogen ionizing
radiation.  Optically thin clouds generally occupy the lower left
portion of the distance -- density plane below a line with slope $\rm{d
\log}(R)/ \rm{d\log}(n)$ $\sim\/-$2/3 and intersecting the lower right
hand corner.  The temperature of this gas (see Fig.~1) ranges from
20,000~K to several million K.  Emission lines presented in this paper
are not emitted by gas with T$_e$~\gtsim\/ 10$^5$~K (X-ray lines {\em
are}).  These hot clouds are generally truncated by the plane parallel
condition discussed earlier.  Optically thick clouds are in the upper
right portion of the figures, and the gas generally has nebular
temperatures, except the extreme upper right corner where the electron
temperatures can be as low as a few hundred Kelvin and the gas is
mostly molecular.  Generally the optically thick clouds were stopped
because the electron temperature fell below 3000~K ($\S$\/2.1).  The
total hydrogen column density restriction mentioned in $\S$\/2.1
applies to a small number of high density optically thick clouds in a
triangular region bounded by a distance of 1~pc on the the topside and
a line roughly parallel to $\log~U \sim~-1.0$ on the leftside.  The
clouds in this region do not significantly contribute to the narrow
forbidden line spectrum.

Different types of lines can be seen in Figure~1 to be emitted from
clouds with a range in gas densities and distances.  Recombination
lines, such as $H\beta$ and $L\alpha$, generally form broad planes of
maximum equivalent width whereas the forbidden lines, such as [\nev\/],
are either narrow ridges or peaked islands ([\oii\/] $\lambda$7325).
Figure~1b shows the differences in the nebular and auroral lines of
[\oii\/] and [\oiii\/].  Notice that the auroral features are formed
optimally at higher densities, because of their higher critical
density.  Figure~1c shows the contour plots of a few of the stronger
UV-lines.  Figure~1d shows a few strong [\feii\/] infrared
fine-structure lines from a 16-level model atom.  The model atom has
the same configuration as that shown in Thompson (1995) and uses the
collision data of Pradhan \& Zhang (1993) and Einstein As from Quinet,
Le Dourneuf \& Zeippen (1996).  Verner et~al.\ (1997) compare the
differences between the 16-level model atom and a much larger
definitive 371-level calculation.  Preliminary comparisons show that
predicted line intensities of the 16-level atom differ typically by a
few percent and at maximum by 10\% from the larger 371-level
calculation.

To illustrate the problems encountered when trying to use conventional
diagnostic line intensity ratios for an ensemble of clouds with a wide
range of physical conditions, we momentarily assume that each emission
line in the total spectrum is produced with the maximum possible
equivalent width found in the figures.  We then use this spectrum to
reproduce the standard density and temperature diagnostic line ratios.
This is the simplest way to document the effects of a distribution of
clouds.  Three density indicators are given by Osterbrock (1989). These
include the [\suii\/] $\lambda$6716/$\lambda$6731 and [\oii\/]
$\lambda$3729/$\lambda$3726 doublets and the [\ciii\/]
$\lambda$1907/~\ciii] $\lambda$1909 ratio.  For [\feii], Pradhan \&
Zhang (1993) use the 1.534~$\mu$m/1.644~$\mu$m ratio as a density
indicator.  Assuming a temperature of 10,000~K, the [\oii\/] ratio we
predict gives an n$_e$ of 300~cm$^{-3}$, for the predicted [\suii\/]
doublet the density is 700~cm$^{-3}$, for the [\feii\/] ratio n$_e$ is
20,000~cm$^{-3}$, and the [\ciii\/]/~\ciii\/] ratio gives a density of
nearly 300,000~cm$^{-3}$.  The [\oiii\/]
($\lambda$4959~+~$\lambda$5007)/$\lambda$4363 temperature -- density
diagnostic using the peak emission of Figure~1 and an electron
temperature of 10,000~K gives an electron density on the order of a few
million~cm$^{-3}$.  The point of this exercise is clear: {\em line
diagnostics fail when the gas has a wide range of physical
conditions}.  As Figure~1 shows, many of these lines form at different
densities and temperatures.  However, by using the diagnostic line
ratios, one assumes that the lines are formed in gas of similar
physical conditions, which is obviously not the case in Figure~1, and
is likely not the case in nature.  We will in a later section integrate
the ``visibility functions'' of Figure~1 to calculate an emission
spectrum by assuming that the observed spectrum is the result of an
ensemble of clouds with differing properties.

\subsubsection{A dusty NLR}
In the previous section we have assumed that the narrow line region is
a dust free environment.  We now recompute the grid assuming that the
clouds are dusty.  Grains will not survive in an environment with high
photoionizing flux due to the various sublimation processes discussed
by Laor \& Draine (1993).  Therefore, after the manner of Netzer \&
Laor (1993) we vary the abundances as a function of distance from the
ionizing source, in order to mimic these destruction processes.  At
small distances from the central source we assume the abundances are
solar, as given in the previous section.  At a distance of approximately
$10^{16.9}$~cm from the central source (for the ionizing luminosity
given above), Orion type graphite grains (Baldwin et~al.\ 1991) are
just at their sublimation temperature ($\sim$1750~K) at the cloud
face.  At distances larger than this we include graphite grains within
the calculation of the single cloud and deplete carbon such that its
abundance is typical of an \hii\/ region (see below and Baldwin et~al.
1996).  Orion type silicate grains sublimate at $\sim$1400~K,
corresponding to a distance of $10^{17.6}$~cm. Beyond this radius the
abundances are set to values which approximate those found in the Orion
nebula:

\noindent
H :1.00E+00  He:9.50E-02  C :3.00E-04  N :7.00E-05  O :4.00E-04  
Ne:6.00E-05  Na:3.00E-07  Mg:3.00E-06  Al:2.00E-07  Si:4.00E-06  
S :1.00E-05  Cl:1.00E-07  Ar:3.00E-06  Ca:2.00E-08  Fe:3.00E-06 
Ni:1.00E-07 

\noindent
These abundances are based upon the results of several recent studies
of the Orion nebula (Baldwin et~al.\ 1991, Rubin et~al.\ 1991, 1992,
and Osterbrock, Tran \& Veilleux 1992).  Although this discontinuous
``turning on'' of grains is unphysical and our choices of grain
composition and gas depletions are specific, it serves as an example to
simulate the presence of dust in the narrow line region.  The grain
physics used here is described by Baldwin et~al.\ (1991).

Figure~2 shows the same contour plots as Figure~1 except that the
individual models include the grains as specified above.  The
discontinuous grain sublimation distances  are easily seen in the figure
as sharp discontinuities in the contour plots at $\log R = 16.9$ and
$17.6$~cm.

Comparisons of Figures~1 and 2 show four major effects of grains:  (1)
the depletion of refractory elements; (2) the weakening of emission
lines due to absorption of the incident continuum by dust at large
$U(H)$; (3) the photoelectric heating of the gas by grains; and (4) in
the case of resonance lines such as $L\alpha$ $\lambda$1216 and \civ\/
$\lambda$1549, line destruction by grains.

Higher ionization lines, such as [\oiii\/], [\neiii\/], and [\nev\/],
tend to be emitted less efficiently than in the dust-free models, with
the equivalent width falling by as much as a factor of 2.  The
weakening of these lines is caused by the absorption of the incident
continuum by the dust.  The lower ionization lines ([\suii\/],
[\oii\/], [\nii\/], [\oi\/] etc.) actually have their peak emission
increase, though at lower ionization parameters, due to the
photoelectric heating of the gas by the grains.  Resonance line
destruction is readily apparent in Figure~2c.  The heavy depletion of
iron is also apparent in Figure~2d, wherein the peak equivalent width
drops by nearly a factor of ten.  This last point will be discussed
more fully in a later section.

\section{Integrated line emission}
The equivalent widths shown in Figures~1 and 2 reflect the efficiency
with which the clouds reprocess the incident continuum.  It is clear
that different lines are formed optimally in different places in the
gas density -- distance plane.  If the distribution function of clouds
in this plane is known, then the total line luminosity emitted by a set
of clouds will be given by
\begin{equation} L_{line} \propto
\int\!\!\int\ r^{2}F(r,n)\,\psi(r,n)\,dn\,dr,\end{equation} 
where $F(r,n)$ is the emission line flux of a single cloud at radius
$r$, gas density $n$, and $\psi(r,n)$ is the cloud distribution
function, which is not necessarily analytical.

The remaining question is the spatial distribution of the NLR clouds.
Many type 2 Seyferts have resolved NLR's, often with complex structures
that are associated with linear or double-lobed radio sources  (Bower
et~al.\ 1995; Pogge \& De Robertis 1995; Capetti et~al.\ 1996; Cooke
et~al.\ 1997), whereas the structure of the NLR in type 1 Seyferts is
generally more compact and  axisymmetric, at least in projection
(Schmitt \& Kinney 1996; Nelson et al. 1996).  Since it is not clear
how the emitting gas is distributed, in {\em all} its dimensions, we
will assume that the distribution function $\psi(r,n)$ can be
approximated by $\psi(r,n)~\propto~f(r)g(n)$, where $f(r)$ and $g(n)$
are the cloud covering fractions with distance and gas density,
respectively (see Baldwin et~al.\ 1995).  For simplicity we assume
simple power laws:
\begin{equation}
f(r) \propto r^{\gamma} ~,~ g(n) \propto n^{\beta}.
\end{equation}
The power law indices, $\gamma$ and $\beta$, will be the only free
parameters of our predicted integrated spectra.  In the case where both
weighting functions are proportional to the inverse of $n$ or $r$, then
the line equivalent widths of the clouds are equally weighted (in log
space) across the gas density -- distance plane.

In the integrations computed below we assumed certain integration
limits in the gas density -- distance plane in order to exclude gas
that may not be present in the NLR.  We exclude gas with densities
$\log n(H) > 8$, the rough maximum critical density of the optical
forbidden lines observed.  By including gas with $\log n(H) \sim 9$,
the line ratios do not change significantly except for those
integrations that are weighted to higher densities (flatter $g(n)$). If
present, the contribution of high density gas in the inner regions of
the NLR cannot be significant or else the narrow line spectrum would
appear vastly different (forbidden lines would be too weak relative to
the hydrogen lines).  We exclude gas that is closer than 0.1~pc to the
ionizing source.  This distance corresponds to the grain sublimation
point described in $\S$2.3. Integration limits were also included in
ionization parameter space in order to exclude gas that is far over
(under) ionized and does not contribute to the observed emission.
These limits effectively exclude gas from the broad emission line
region and very high density gas hundreds of parsecs away from the
central engine.

\subsection{Diagnostic line ratio diagrams}
Integrations described above have been computed and plotted after the
manner of Baldwin, Phillips \& Terlevich (1981; hereafter BPT).
Intensity ratios for the integrated model spectra are plotted in this
manner in Figure~3.  Two types of lines are shown in the figure.  Solid
lines represent dusty models with depleted refractory elements typical
of \hii\/ region type abundances (hereafter ``dusty''). The dashed
lines represent dust-free clouds with solar metallicities (hereafter
``dust-free'').  Different values of $g(n)$ (cf. Eq.~[2]) are
represented by different lines, with the power of $f(r)$ ranging from $-2.0 \leq
\gamma \leq 1.0$ in 0.25 increments along each line with the direction
of more negative $f(r)$ shown by the arrows.  For the dusty simulations
(solid lines) three values of the $g(n)$ index are shown ($\beta$ =
$-$1.0, $-$1.6, $-$1.8), and the trend of more negative $\beta$ are
indicated by the appropriate arrows in the figures.  For the dust-free
(dashed lines) simulations the $g(n)$ index are $\beta$ = $-$1.0,
$-$1.4, $-$1.8, similarly the trends for the more negative index are
indicated.  The ``S'' and ``L'' indicate in the figures the average
values of the line ratios for Seyfert, taken from Ferland \& Osterbrock
(1986), and LINER type spectra taken from Netzer (1990), with the
exceptions of \heii $\lambda$4686, [\oiii\/] $\lambda$4363, from Ho,
Filippenko \& Sargent (1995) for the LINER galaxy M81.  The observed
range in [\oii\/] $\lambda$7325 and [\suiii] $\lambda\lambda$9069,9532
strengths were taken from Dopita \& Sutherland (1995).

To make it possible to follow particular models from panel to panel, we
have marked two fiducial models in Figure~3. The stars represent a
dusty model (solid lines) with  $f(r) \propto r^{-1.5}$ and $g(n)
\propto n^{-1.6}$, while the squares represent a dust-free model
(dashed lines) with  $f(r) \propto r^{-1.25}$ and $g(n) \propto
n^{-1.4}$. These particular models were chosen because they
approximately reproduce the [\oii\/] $\lambda$3727/[\oiii\/]
$\lambda$5007 and [\oiii\/] $\lambda$5007/H$\beta$ line ratios in the average Seyfert 2 spectrum.

Interestingly, solar abundance, grainless calculations are not excluded
by the line ratios presented in Figure~3.  The dashed, dust-free, lines
in Figure~3 tend to converge in the area of Seyfert observations.  The
dusty integrations tend to be shifted toward and into the LINER region
of the diagrams, that is with higher [\nii\/], [\suii\/], and [\oi\/]
relative to H$\alpha$ and higher [\oii\/] relative to [\oiii\/].

Table~1 compares the fiducial models represented by the star and square
in Figure~3 with the observed average Seyfert 2 spectra from Ferland \&
Osterbrock (1986), Dopita \& Sutherland (1995), Thompson (1995), and
Villar-Mart\'{\i}n \& Binette (1997).  The UV/optical line ratios given
by Ferland \& Osterbrock, likely suffer from inaccurate absolute
optical spectral flux calibrations coupled with substantially different
aperture sizes between the IUE UV spectra and ground based optical
spectra coupled with the extendedness of the NLR, making many of these
UV/optical narrow line ratios highly uncertain.  The table also lists
the observed range of equivalent width of H$\beta$ taken from Netzer
(1990) and Binette, Fosbury, \& Parker (1993).

In column 4 of Table~1 we give the dereddened UV/optical HST/FOS
spectrum of the Seyfert 2 galaxy NGC~3393, from Cooke et~al.\ (1997).
The Cooke et~al.\ observations were taken at two overlapping but
significantly different aperture positions within a complicated
two-dimensional line-emitting structure; spectra covering 1100~\AA\/ --
2300~\AA\/ were taken at one position while the range 2300~\AA\/ --
6800~\AA\/ was covered at the other position.  Because of the
unintended positional offset it is impossible to unambiguously join
together the two spectral regions.  We have normalized the two sections
of the NGC~3393 spectrum so that the UV/optical \heii\/ ratios would
agree with case B predictions.  This does not guarantee consistency
amongst the other emission lines, but it is not clear that the
uncertainty due in this case to a 0.4 arcsec offset between two 1
arcsec diameter apertures is any worse than the errors incurred through
the comparison of IUE to ground based data, whereas the signal:noise in
the NGC~3393 spectra is much higher than in the earlier data.

Observations by Oliva et~al.\ (1994) of the Seyfert type 2 Circinus
galaxy are given in column 5 of Table~1.  These observations include
the ground-based optical and infrared spectra of many high ionization
``coronal lines'' (ionization potentials greater than 100~eV).

Columns 6 and 7 of Table~1 show the results of the predicted integrated
line spectrum for the dust-free and the dusty fiducial points.  Recall
that the integration parameters for the two fiducial models were chosen
to simultaneously reproduce the [\oiii\/]/[\oii\/] and
[\oiii\/]/H$\beta$ emission line ratios of the average Seyfert galaxy,
and with this in mind, the integrated spectrum is representative of the
observed values.

\subsubsection{UV lines}
Compared to the mean Seyfert 2 spectrum the UV lines, L$\alpha$,
\civ\/, and \ciii\/], are generally underpredicted by the solar,
dust-free integration shown in column 5 of Table~1.  In column 6 of the
table, the dusty integration underpredicts L$\alpha$, and \civ\/ even
worse than the dust-free case (generally because of resonance line
destruction by dust grains).  The exceptions being that \ciii\/]
$\lambda$1909 line, is within the observed range, and \mgii\/ is on the
low end of the range.  The predictions of the weaker UV emission lines
observed in NGC~3393, not included in the mean Seyfert 2 spectrum of
Table~1, are generally consistent with the observations.  The
execptions \nv\/ and \heii\/ will be elaborated upon in the next
section.

The integrations of the dust-free clouds predict
L$\alpha$/H$\beta$~$\sim$\/28, which is close to the high density case
B limit ($\sim$34), while the observations compiled by Ferland \&
Osterbrock (1986) suggest much larger ratios (30 -- 70).  Our
calculations find a maximum L$\alpha$/H$\beta$~$\sim$\/100 occuring at
$\log n(H)~$=~9.75~cm$^{-3}$.  L$\alpha$/H$\beta$ significantly larger
than $\sim$34 implies the existance of a substantial population of high
density clouds; more specifically 40 $\leq$ L$\alpha$/H$\beta$ $\leq$
70 occurs for 5.5 $\leq \log n(H) \leq$ 8.5.  However, the spectrum
resulting from integrations that heavily weight this dense gas would
not be consistent with the strong optical forbidden line strengths
relative to H$\beta$. 

Our predicted hydrogen line spectrum is in better agreement with the
HST data of Cooke et~al.\ (1997, see Table~1).  The corrected observed
L$\alpha$/H$\beta$~$\sim$\/19 is inconsistent with the compilations of
Ferland \& Osterbrock (1986), with the caveat that this observed ratio
is somewhat uncertain.  L$\alpha$/H$\beta$ ratios smaller than 23 are
not possible, unless L$\alpha$ is weak either due to very dense
($>$~10$^{11}$~cm$^{-3}$) or dusty gas.  The NGC~3393 \mgii\//H$\beta$
ratio is also consistent with the dusty simulations.  HST observations of
the narrow lines of several high luminosity AGN (Wills et~al.\ 1993)
also yield ratios that are more consistent with the dusty simulations.
Clearly, uniform, long slit, high signal-to-noise, UV to infrared
spectra are required to understand the narrow emission line spectra of
AGNs.  Future observations using the STIS on HST will accomplish a good
part of this for the first time.

\subsubsection{Optical lines}
The predicted narrow optical forbidden lines are generally within the
observed ranges of the mean spectrum with the exceptions of [\neiii\/]
and [\nev\/] both being slightly underpredicted by $\sim$50\% and
$\sim$20\%, respectively, indicating possible differences in
abundances.  However, these lines are predicted to be stronger relative
to H$\beta$ than observed in NGC~3393.  The [\suii\/], [\suiii\/] lines
are within the observed ranges of the mean Seyfert~2 spectrum.

Both integrations of the predicted nitrogen line [\ni\/], are consistent
with the observed mean Seyfert 2 range, although the [\nii\/] line is
predicted to be lower that observed by nearly half.  Comparisons
between both integrations and NGC~3393 of the nitrogen emission lines
([\ni\/], [\nii\/], and \nv\/) relative to H$\beta$ indicate that
nitrogen may be enhanced in that object relative to solar by a factor
of $\sim$2.

The strength of the \heii\/ $\lambda$4686 permitted line relative to
H$\beta$ is often used to constrain the ionizing continuum.  A stronger
\heii\/ $\lambda$4686 line (relative to H$\beta$) is indicative of more
He$^+$ ionizing photons.  Simple photoionization models often
underpredict the line while doing well with the rest of the optical
spectrum (Ferland \& Osterbrock 1986).  Figure~3 and Table~1 show that
the integrated model can predict this line within the observed range.
Our fiducial integrations given in columns 5 and 6 of the table
overpredict the entire \heii\/ spectrum relative to H$\beta$ when
compared to NGC~3393, indicative that NGC~3393 has a slightly softer
continuum than the one assumed here ($\S\/$2.1).

The predictions of the relative strengths of the optical iron coronal
lines, [\fevii\/], [\fex\/], [\fexi\/], and [\fexiv\/] to H$\beta$ are
in accord with the observations of the Circinus galaxy (Oliva
et~al.\ 1994) for the dust-free integrations.  The dusty integrations
underpredict these lines by factors of $\sim$ 5 -- 10. However, Oliva
et al.\ (1994), Morisset \& P\'{e}quignot (1996), and Ferguson et
al.\ (1997) discuss the apparent large uncertainties in the collision
strengths of [\fex\/] $\lambda 6375$, [\fexi\/] $\lambda 7892$, and
[\fexiv\/] $\lambda 5303$. Given these uncertainties we set the
collision strengths for these transitions equal to 1 prior to running
these simulations (note: Ferguson et al.\ used the collision strengths
from the literature). These lines will not be useful probes of the high
ionization narrow line region until this issue is resolved.

The weakness of the optical doublet [\caii\/] $\lambda\lambda$7291,7324
has been cited as evidence for the presence of dust grains with the NLR
(Ferland 1993; and Kingdon, Ferland \& Feibelman, 1995).  Table~1
indicates that the computations assuming solar abundances overpredict
the [\caii\/] $\lambda$7291 line by factors of 3 -- 160, with the dusty
calculations on the very low end of the observations.  This suggests
that calcium is actually depleted by factors up to 160 compared to its
solar abundance.

\subsubsection{Infrared lines}
The infrared forbidden [\feii\/] lines from the integrations described
above are compared with the observations of NGC~4151 made by Thompson
(1995).  Because Thompson's observations included P$\beta$, we can
reference the [\feii\/] to H$\beta$, as in Table~1, by assuming case B
conditions for hydrogen.  This is a valid assumption since
P$\beta$/H$\beta$ predicted by our integrations is very nearly the case
B ratio.  Table~1 shows that the predicted intensities of the
``dust-free'' integration (column 3) are clearly too strong to match
the observations, with some lines overpredicted by a factor of 10.
Even for the ``dusty'' integration (column 4), where the iron
abundance is depleted by a factor of ten, relative to solar, the
strongest [\feii\/] lines are too strong by a factor of 2 compared with
H$\beta$.  This evidence suggests that iron may be depleted by factors
up to 20 in NGC~4151.

Simpson et~al.\ (1996) and Veilleux et~al.\ (1997) present IR
spectroscopy of a number of Seyfert~2 galaxies.  Simpson et~al.\ found
[\feii\/] 1.257 $\mu$m/P$\beta$ $\approx$ 1.21~($\sigma~\approx$~0.81),
and Veilleux et al.\ found [\feii\/] 1.257 $\mu$m/P$\beta$ $\approx$
1.24~($\sigma~\approx$~0.90).  As in NGC~4151 these observed values are
better matched to the predicted ratio from the dusty integration in
Table~1 ($\approx$~1.4). The dust-free integration predicts a ratio of
$\approx$~7, much like the ratios arising in the shocked (grain free)
environments of supernova remnants, plotted by both Simpson et~al.\ and
Veilleux et~al.

Although the [\feii\/] line ratios are too strong compared with the
hydrogen lines, indicating an abundance difference, the relative
[\feii\/] spectrum is fairly consistent with the observations.
Thompson (1995) discusses at length the probable difficulties with the
observation of weaker lines, such as 1.279~$\mu$m and 1.677~$\mu$m.

Table~1 also includes predictions of the relative strength of the
far-IR [\feii\/] 25.98$\mu$m line, the lowest transition within the
$^6$D ground term.  This line is predicted to be quite strong in the
solar integration (4 times P$\beta$). Recall that the simulations
presented here excluded gas with T$_e <$ 3000~K, and emission from
these low excitation lines might still be important at lower electron
temperatures. In fact, we find that this line and others that arise
within the ground term will have substantial contributions from
AGN-heated photodissociation regions, if present.  None of the $\sim$
1~$\mu$m [\feii\/] lines are emissive in gas temperatures below
3000~K.

Our simulations of the infrared coronal lines shown in Table~1 are
generally in accord with the observations of the Circinus galaxy for
the dust-free integration.  The [\sivi\/], [\sivii\/], and [\siix\/]
spectrum is well matched by the fiducial dust-free integrations.  The
dusty models do a poor job at matching the observations, particularly
the [\caviii\/] line.  Our dust-free fiducial model predicts this line
to within a factor of two, whereas the dusty integration underpredicts
this line by over a factor of 30. The predicted strength of the optical
coronal line [\fevii\/] $\lambda$6087 is also consistent with the
proposition that the high ionization lines are emitted in a dust-free
environment.

\section{Line width and critical density}
Filippenko \& Halpern (1984) showed that the line width of optical
forbidden lines is correlated with the critical density of those
lines.  Other observers (Filippenko 1985, Appenzeller \&
\"{O}stereicher 1988, Espey et~al. 1994 and Ho, Filippenko, \& Sargent
1996) have extended this work.  It was realized by Filippenko \&
Halpern that the probable explanation for the observed correlation is
that the lines emit most efficiently near their critical densities and
that the line emission is spatially stratified. Our LOC models are a
quantitative formulation of this idea.  By just glancing at Figure~1,
one observes that the peak emission of the set of lines shown occurs
over a range of gas density and distance.  If the dominant velocity
field is one which diminishes with distance from the nucleus, then we
expect that those lines that are emitted optimally at smaller distances
(larger $n(H)$) to have broader line widths than those that are
optimally emitted farther away (smaller $n(H)$) and qualitatively agree
with the observed correlation.  Consider the contour plots of [\oiii\/]
$\lambda$4363 and [\suii\/] $\lambda$6720 shown in Figure~1. It is
clear by looking at their places of peak emission that [\oiii\/] will
be much broader than [\suii\/], as is observed.

Figure~4 shows the results of a simple calculation of relative line
widths assuming that the density -- distance phase space system of
clouds obey the covering fraction and density indices of the fiducial
square shown in Figure~3.  In lieu of a more realistic velocity field
which takes into account the mass distribution of the bulge in addition
to the supermassive black hole, we assume virial velocities about a
simple point mass of 10$^9$~M$_{\odot}$.   In the figure we show the
full width at half maximum (FWHM) and the full width at 10\% intensity
(FW10\%) for the ions indicated versus the logarithm of the critical
density ($n_{e}^{crit}$).  The slopes of the fitted linear regression
in the figure were computed to be 0.193 (FW10\%) and 0.118 (FWHM),
similar to those observed by Filippenko \& Halpern (1984) and Espey
at~al.  (1994), which range from 0.095 to 0.200.

\section{Application to Resolved NLR's.}
In a number of nearby Seyfert galaxies the NLR is spatially resolved
even on ground-based long-slit and Imaging Fabry Perot spectra. These
NLRs frequently have the form of either ionization cones (cf. Pogge
1988; Tadhunter \& Tsvetanov 1989; Wilson et al. 1993; Robinson
et~al.\ 1994) or of apparent bow-shocks and cocoon shocks around linear
radio sources (Bower et al.  1995; Pogge \& De Robertis 1995; Capetti
et al. 1996; Cooke et al.  1997). Typical scales are 200-400 $h^{-1}$
pc arcsec$^{-1}$, where $h = $H$_o/100$ km s$^{-1}$ Mpc$^{-1}$. With
the imminent commisioning of the HST STIS spectrograph and ground-based
adaptive optics systems, it should be possible to study
spectroscopically a reasonable sample of NLR's with $\sim 20-40$ pc
($\sim 10^{20}$ cm) resolution.

The exact scaling of such dimensions onto the y-axis of Figures 1 and 2
is complicated by the fact that the (probably beamed) ionizing
continuum luminosity seen by these NLRs is not directly measured.
However, simple arguments based on counting recombination photons (cf.
Binette, Fosbury \& Parker 1993) imply that the gas sees an ionizing
luminosity that is on average a factor of 10 larger than what we see.
Therefore, the range in incident flux used in the model grid presented
here is applicable to these resolved NLRs, and it is of interest to
consider the predicted radial dependence of various line intensity
ratios when integrated over the cloud gas density distribution.

Figure~5 shows four different line ratios indicating the gas density,
temperature, and ionization as functions of distance and gas density in
the form of logarithmic contour plots.  The figure also plots the
ratios of the line intensities integrated over clouds of all densities,
as functions of distance from the ionizing source; shown are the
results of two different integrations along the density axis.

\subsection{Density}
The density indicator [\suii\/] $\lambda$6716/$\lambda$6731 is shown in
the top two panels of Figure~5a.  The line ratio has the linear value
of 1.36 in the extreme upper left corner of the plot at ($\log n(H) =
2.0,\log R = 22.0$).  Moving down and to the right along a line of
constant ionization parameter the solid lines are 1 dex increments and
the dashed lines are 0.1 dex steps.  The large flat plain of constant
ratio corresponds to the high density limit or a linear value of
$\sim$~0.44.

The line ratio as a function of distance shows that for the
integrations presented the high density limit is predicted at small
distances.  The solid line (integrated assuming that $g(n) \propto
n^{-1.0}$) remains at that limit for larger distances, whereas the
dashed line (integrated assuming a steeper density distribution, i.e.
$g(n) \propto n^{-1.8}$) turns upwards sooner.  The dashed curve
reaches a peak value nearly equal to the low density limit given by
Osterbrock (1989).  The gas density inferred from the dashed curve
falls off as $R^{-2/3}$, while the density inferred from the solid
curve is pinned at the high-density limit for larger values of R, then
falls off more steeply than $R^{-2/3}$.

\subsection{Temperature}
The temperature indicator [\oiii\/] $\lambda$5007/$\lambda$4363 (note
that $\lambda$4959 is not included) is shown in the bottom two plots of
Figure~5a.  The contour plot has a peak value of 112 in the upper left
hand corner ($\log n(H) = 2.0,\log R = 21.0$).  Following a line of
constant ionization parameter down and to the right, the solid contour
line (which has a backwards ``S'' shape) has the logarithmic value of
2.0, and the dashed lines are 0.2 dex increments.  The outer contour in
the lower right part of the plot has a logarithmic value of $-$1.0 at
($\log n(H) = 10.0,\log R = 18.0$).  The contour plot shows that in the
low density limit ($\log n(H)$ \ltsim 4.5~cm$^{-3}$) the ratio is a good
temperature indicator in that moving perpendicular to the ionization
parameter the ratio changes are roughly independent of density.  At
higher densities the ratio becomes dependent upon both temperature and
density.

The gas density integrated line ratio as a function of distance shows
that for large distances from the ionizing source the ratio is
independent of the weighting of the density axis; i.e.,\ the ratio is in
the low density limit.  Using the analytical expression from Osterbrock
(1989) the ratio can be converted into temperature in the low density
limit.  For the range 20.0~cm $\leq \log R \leq$ 22.0~cm the inferred
temperature has the range 14,000~K $\geq T_e \geq$ 8400~K, falling off
very gradually as approximately $R^{-1/9}$.  At smaller radii the
$\lambda$5007/$\lambda$4363 ratio further decreases, reflecting,
mainly, the drop in $\lambda$5007 due to collisional quenching and the
increase of $\lambda$4363 which continues to emit efficiently up to its
critical density, rather than reflecting an increase in temperature.

The [\nii\/] $\lambda$5755/$\lambda$6584 line ratio is also a useful
temperature indicator.  Wilson, Binette \& Storchi-Bergmann (1997) use
this ratio in concert with the [\oiii\/] $\lambda$5007/$\lambda$4363
ratio as evidence for the presence of matter-bounded clouds in the
NLR.  Defining $R_{[NII]}$ as the [\nii\/] line ratio and $R_{[OIII]}$
as the [\oiii\/] line ratio, our model integrations predict that
$R_{[OIII]}$ versus $R_{[NII]}$ depends little upon the density
weighting factor $g(n)$.  We find a linear-log relationship of the form
\begin{equation}\log R_{[OIII]} = 0.1593\log R_{[NII]} + 0.1932.
\end{equation} The fiducial dust-free integration from Figure~3
($\S\/$3.1) predicts $R_{[OIII]} = 0.0182$ and $R_{[NII]} = 0.0215$,
and the fiducial dusty simulation predicts $R_{[OIII]} = 0.0206$ and
$R_{[NII]} = 0.0229$. These correspond to temperature differences
($T_{[OIII]} - T_{[NII]}$) of 3000~K and 3500~K, respectively. These
predictions lie in the range of observed values (Wilson, Binette \&
Storchi-Bergmann), and do not require the presence of a significant
population of matter-bounded clouds, although we do not exclude this
possibility.

\subsection{Ionization}
The ratio [\oiii\/] $\lambda$5007/(H$\alpha$~+~[\nii\/] $\lambda$6584)
has been used as an ionization indicator and this ratio is shown in the
top left panel of Figure~5b.  The outer contour has the linear value of
1.0 and the dashed lines are $+$0.2 dex steps.  We have chosen to
include both H$\alpha$ and [\nii\/], since many filters include both
lines.  The contour plot of this ratio is relatively flat over a wide
range in density, distance and ionization parameter.  In fact, for a
constant density of $n(H) = 10^4$~cm$^{-3}$, the
[\oiii\/]/(H$\alpha$~+~[\nii\/]) ratio would be constant from $\sim$ 10
-- 50~$L_{43.5}^{1/2}$~pc, even though the ionization parameter
decreases by $\sim$~10$^{1.5}$ over the same range.  A ratio that
behaves in similar manner is [\oiii\/] $\lambda$5007/H$\beta$, and
is shown in the bottom left panel of Figure~5b.  The outer contour has
the linear value of 1.0 and the dashed lines are $+$0.2 dex steps.  The
[\oiii\/]/H$\beta$ ratio is just as flat over a wider range in
ionization parameter as is [\oiii\/]/(H$\alpha$~+~[\nii\/]).  Thus,
while these line ratios are useful in separating \hii\/ region galaxies
from active galaxies (c.f.\ BPT), they are not reliable ionization
parameter indicators.

In the top right panel of Figure~5b the [\oiii\/]
$\lambda$5007/(H$\alpha$~+~[\nii\/] $\lambda$6584) as a function of
distance is shown.  The ratio has the feature of being constant from
$\sim$ 3 -- 300~$L_{43.5}^{1/2}$~pc for the integration assuming a
steep distribution ($g(n) \propto n^{-1.8}$; dashed line) of clouds
along the density axis.  The solid line ($g(n) \propto n^{-1.0}$) is
flat over a much smaller distance range, because in this integration
emission from higher density gas is being included.  The sharp falloff
of the ratio at large distances ($R \geq 300~L_{43.5}^{1/2}$~pc) occurs
because the [\oiii\/] $\lambda$5007 emissivity is dropping very quickly
with distance at the lowest gas density (10$^2$~cm$^{-3}$) considered
here, since the gas is underionized and can not produce the line
efficiently, while the [\nii\/] $\lambda$6584 line is simultaneously
increasing in strength (see Figure~1).  The bottom right panel of
Figure~5b shows the integrated [\oiii\/]/H$\beta$ ratio as a function
of distance; this ratio behaves similarly to the
[\oiii\/]/(H$\alpha$~+~[\nii\/]) ratio.  Both of these line ratios are
observed to be relatively constant over factors of 3 -- 4 in distance
in several Seyfert 2 galaxies.  (Robinson et~al.\ 1994; Capetti
et~al.\ 1996).

Figure~6a shows the [\oiii\/] $\lambda$5007/[\oii\/] $\lambda$3727 line
ratio as a contour plot.  In the figure, the outer contour has a linear
value of 1.0, and the dashed lines are $+$0.2 dex steps.  At low
densities, the contours are closely spaced and parallel with ionization
parameter, indicating that this line ratio is a good ionization
indicator.  At densities $\sim$~10$^{4.5}$~cm$^{-3}$, [\oii\/] becomes
collisionally quenched, while [\oiii\/] does not, thus the ratio becomes
dependent upon density as well.

In Figure~6b we show the predicted [\oiii\/]/[\oii\/] line ratio as a
function of radius for 4 models compared with the observations of
Robinson et~al.\ (1994) for NGC~4151.  The data have been placed on the
distance axis by assuming that 1\arcsecpoint\/0 $\approx$ 100~pc, as
used by Robinson et~al.  The model integrations have been rescaled in
distance by factors of 4, corresponding to a 16-fold increase in the
ionizing luminosity incident upon the clouds.  The observed strengths
of these and a number of other emission lines in NGC~4151 suggest that
the ionizing luminosity incident upon the narrow-line clouds may be 10
-- 20 times the observed one (Penston et~al.\ 1990; Robinson et~al.),
though the magnitude of the photon deficit is very uncertain (see also
Schulz \& Komossa 1993).  Since the ionizing luminosity employed here
happens to match that of NGC~4151 emitted toward Earth (Robinson et
al.), our rescaling is consistent with the continuum beaming inferred
from the observations.  The predicted [\oiii\/]/[\oii\/] ratio shown in
Figure~6b did not include emission from matter-bounded clouds;
including such clouds would tend to enhance the [\oiii\/]/[\oii\/]
ratio where it is small in Figure~6b.

Robinson et~al.\ (1994) include several other line ratios as functions
of projected distance for NGC~4151.  These include [\nev\/]/[\neiii\/]
and [\oiii\/]/[\nii\/] which behave similarly to [\oiii\/]/[\oii\/]
shown in Figure~6b.  Also observed are [\nii\/]/H$\alpha$,
[\oi\/]/H$\alpha$, [\oii\/]/H$\alpha$, and [\suii\/]/H$\alpha$.  Our
simulations qualitatively match the general trends in these line
ratios.

\section{Is the NLR Dusty?}
The presence of dust in the NLR is a vital clue to the origin of the
gas.  For instance, if shocks are prevalent in the emitting region,
then grains will not survive (Donahue \& Voit 1993).  Both the
dust-free and dusty simulations reproduce the average narrow emission
line ratios shown in Figure~3.  However, other lines do indicate
substantial depletions relative to solar.  The absence of an observed
[\caii\/] $\lambda$7291 line has led to speculation that the NLR is a
dusty place (Ferland 1993; and Kingdon, Ferland \& Feibelman, 1995).
Recent observations of the [\caii\/] line (Villar-Mart\'{\i}n \&
Binette 1997) show that the line is much weaker than the solar
integrations predict (Table~1), supporting Villar-Mart\'{\i}n \&
Binette conclusion that Ca is sharply depleted.  The simulations
presented here indicate that Ca is depleted relative to solar by
factors of 3 -- 160 based on the range of observed line strengths when
compared to the dust-free simulations in Table~1.  The results of the
simulations for the infrared [\feii\/] lines, as discussed in
$\S$3.1.3, indicate that Fe is also underabundant relative to solar by
factors of 10 -- 20 in NGC~4151.  Comparisons of the compilations of
the [\feii\/] 1.257$\mu$m/P$\beta$ line ratio by Simpson et~al.\ (1996)
and Veilleux et~al.\ (1997) with the simulations provide further
evidence that the NLR is a dusty environment.  A similar conclusion was
reached by Simpson et~al.

Two line ratios, L$\alpha$/H$\beta$ and \mgii\//H$\beta$, observed in
NGC~3393 and discussed in $\S$\/3.1.1, also suggest that dust exists in
the emitting gas.  However, the uncertainty in the correction for the
unintended positional error in the observations, makes the conclusion
from L$\alpha$/H$\beta$ uncertain.  There is no positional error for
the \mgii\//H$\beta$ ratio, and this ratio makes a stronger statement
supporting the presence of dust in NGC~3393.

Evidence that a portion of the emitting gas is dust free is apparent in
the observations of strong coronal lines in many AGN.  Oliva et al.
(1994) have observed a very strong [\caviii\/] 2.32~$\mu$m infrared
line and strong optical [\fevii\/] $\lambda$6087, [\fex\/]
$\lambda$6374, [\fexi\/] $\lambda$7892, and [\fexiv\/] $\lambda$5303
lines, indicating that the gas is dust-free.  Simulations of coronal
line emitting gas by Korista \& Ferland (1989), Oliva et~al.\ (1994)
Ferguson, Korista, \& Ferland (1997), and the results of those shown in
Table~1 indicate that these elements are not depleted.

Taking the observations at face value, we are presented with a
conundrum:  the low ionization UV \mgii\/, optical [\caii\/], and
infrared [\feii\/] lines clearly indicate that dust is present, but the
strong coronal lines suggest a dust-free environment.  One possible
reconciliation is to postulate that there is dust in the NLR, but only
in the neutral or partially neutral portions of the clouds.  In the
simulations presented in this paper we have assumed that the entire
cloud is either dust-free or dusty.  It is not unreasonable to imagine
that the highly ionized portion of NLR clouds to be dust free, perhaps
because the grains have been exposed to a strong UV continuum for quite
some time, and that in more neutral regions grains are free to exist.
Another possibility is that high ionization clouds have been processed
through shocks, but those which emit more neutral ions have not.  Both
of these possiblities argue for a very complex NLR environment, where
the emitting gas has a variety of origins.

\section{Summary}
We have shown that the AGN narrow emission line spectra, from [\oi\/]
to [\siix\/], can largely be reproduced by assuming that the emitting
region consists of clouds with a wide range of gas densities and
distances from the ionizing source. Here we chose simple power-law
cloud distribution functions and the appropriate power law indices
($\beta$ and $\gamma$) whose resulting spectrum matched the observed
[\oiii\/]/[\oii\/] and [\oiii\/]/H$\beta$ ratios. While smooth power
law distributions in cloud properties may be too simple, its success in
predicting the observed spectrum points to the strength of the
``locally optimally emitting cloud'' scenario. The LOC model
integrations result in the following:

\noindent
(1) The classical BPT line ratios for AGN narrow emission lines are
reproduced (Figure~3).

\noindent
(2) The predicted H$\beta$ equivalent width, assuming full source
coverage and an isotropic continuum, is more than sufficient to account
for the observed emission in most Seyfert galaxies (Table 1). An
anisotropic continuum will strengthen the predicted line equivalent
widths and allow for smaller covering fractions. The predicted
[\oiii\/]/[\oii\/] ratio as a function of radius implies that the
narrow line clouds see a continuum that is an order of magnitude more
luminous than we observe in NGC~4151.

\noindent
(3) The LOC integration simultaneously reproduces the low [\suii\/]
$\lambda$6716/$\lambda$6731 density and the high [\oiii\/]
$\lambda$5007/$\lambda$4363 temperature (cf. Figure 3b), resolving a long
standing problem.

\noindent
(4) The observed line width -- critical density correlation for the
optical forbidden lines can be reproduced.  A natural consequence of an
ensemble of clouds in the gas density -- distance plane is that lines
with higher critical densities are formed at smaller distances (cf.
Figure~1).  We have shown that if we assume that this ensemble of
emitting clouds is placed in a gravitational potential, then the line
width -- critical density correlation is very similar to the observed
will naturally result.

\noindent 
(5) The presence of a large density range at each radius can change the
values of the classical density and ionization-level indicators to at
least more closely mimic the observed radial gradients in these
intensity ratios. 

\noindent 
(6) Low ionization lines show clear evidence of depletion (Figure~3 and
Table~1), but high ionization regions do not.  It is possible that dust
grains are selectively destroyed in high ionization regions by shocks
or sublimation and that the neutral regions are undisturbed.  This
argues for a variety of NLR cloud origins.

\acknowledgments
We thank Andrew Cooke for agreeing to our use of the NGC~3393 data.  We
acknowledge suport from NASA and NSF through NAG-3223 and AST93-19034,
and STSCI for GO-06006.02.  We also thank the University of Kentucky
Center for Computational Science and its director John Connolly for
access to the Exemplar parallel supercomputer.

%
%
\newpage
\begin{center}
{\bf Figure Captions}
\end{center}

\noindent 
Fig.~1.--- {Contours of constant logarithmic line equivalent widths as
a function of $\log R$ and $\log n(H)$ for the 23 ions indicated,
referenced to the incident continuum at 4860~\AA\/.  The bold lines
represent 1 dex increments and the dotted lines are 0.2 dex steps.  The
triangle is the peak of the equivalent width distribution and the
contours decrease downward to the outer value of 1~~\AA\/. The reader
will sometimes find it convenient to view the contour plots {\em along}
the ridge at large inclination angle to the sheet of paper.  The upper
right hand plot on the first panel of the figure is the log T$_e$ of
the front face of the cloud.  The temperature decreases from 10$^7$~K
in the lower left hand corner of the plot to 10$^3$~K in the upper
right hand corner.  Again, the bold lines represent 1 dex increments
and the dotted lines are 0.2 dex steps.}

\noindent 
Fig.~2.--- {Same as Fig.~1, except for the dusty abundance set described
$\S$2.2.2.}

\noindent
Fig.~3.--- {Line diagnostic diagrams after the manner of Baldwin,
Phillips \& Terlevich (1981).  Different line types represent different
abundance types, dashed are solar or dust-free, and solid lines are the
dusty abundance set described in the text.  Different lines represent
different families of the density weighting index, $\beta$ with
different covering fraction indices, $\gamma$ along each line with the
arrows indicating the direction of the more negative index.  The star
represents the same model integration of dusty abundances with indices
$f(r) \propto r^{-1.5}$ and $g(n) \propto n^{-1.6}$.  The square
represents the same model integration with solar abundances with
indices $f(r) \propto r^{-1.25}$ and $g(n) \propto n^{-1.4}$.  The
large ``S'' and ``L'' symbols refer to the observations of the mean
Seyfert type 2 objects and LINER type objects, respectively, as
discussed in $\S$3.1.}

\noindent 
Fig.~4.--- {Line width versus critical density of the square point
shown in Fig.~3.  The solid circles are the full width at 10\%
intensity with a fitted slope of 0.193.  The open squares are FWHM with
a fitted slope of 0.118. The observed range in the slope is from 0.095
to 0.200.}

\noindent
Fig.~5.--- {Line ratio diagnostic diagrams of gas density, temperature,
and ionization shown in two different forms.  The left hand column of
plots in the figure are logarithmic contour diagrams of the ratios
indicated taken from the simulations shown in Figure~1.  The right hand
column of plots in Figure~5 are graphs of the indicated line ratio
versus distance from the central source for two different integrations
as discussed in $\S$3, but integrated in the density dimension only.
The solid lines are integrations assuming that $g(n) \propto n^{-1.0}$,
and the dashed lines assume $g(n) \propto n^{-1.8}$.  Each line ratio
is shown on a different scale and we will discuss each in turn in the
text in $\S$5.}

\newpage
\noindent
Fig.~6.--- {The predicted [\oiii\/] $\lambda$5007/[\oii\/]
$\lambda$3727 ratio shown in two forms: (a) is a logarithmic contour diagram
of the ratio taken from the simulations shown in Figure~1.  The outer
contour has a linear value of 1, and the solid lines are 1.0 dex
increments with the dashed lines being 0.2 dex steps.  (b) The line
ratio as function of distance from the ionizing source.  The lower
solid line is an integration in density assuming $g(n) \propto
n^{-1.0}$, and the upper solid line assumes $g(n) \propto n^{-1.8}$.
The dashed line is the fiducial integration with $g(n) \propto
n^{-1.4}$ for the dust-free simulation.  The dotted line is the
fiducial integration with $g(n) \propto n^{-1.6}$ for the dusty
simulation.  All integrations have been rescaled by factors of 4 in
distance corresponding to 16 in luminosity (see $\S$5.3).  The filled
circles are the observations of NGC~4151 by Robinson et~al.\ (1994),
assuming 1\arcsecpoint\/0 $\approx$ 100~pc.}

%
\newpage
\begin{table}[h]
\begin{center}
\begin{tabular}{ccccccc}
\multicolumn{7}{c}{\sc TABLE~1} 
\\[0.1cm]
\multicolumn{7}{c}{\sc Observed versus integrated spectra}
\\[0.2cm]
\hline
\hline
%
\multicolumn{1}{c}{Line}
&\multicolumn{1}{c}{Mean Sy 2$^a$}
&\multicolumn{1}{c}{Reference$^b$}
&\multicolumn{1}{c}{NGC~3393$^c$}
&\multicolumn{1}{c}{Circinus$^d$}
&\multicolumn{1}{c}{Solar$^e$}
&\multicolumn{1}{c}{dusty$^f$}\\
\multicolumn{1}{c}{(1)} & \multicolumn{1}{c}{(2)} & 
\multicolumn{1}{c}{(3)} & \multicolumn{1}{c}{(4)} &
\multicolumn{1}{c}{(5)} & \multicolumn{1}{c}{(6)} & 
\multicolumn{1}{c}{(7)}
\\[0.05cm]
\hline

W$_{\lambda}$(H$\beta$)$^g$ & 5 -- 30 & N &  -- & -- & 152 & 61 \\

W$_{\lambda}$(H$\beta$)$^g$ & 15 -- 100 & BFP & --  & -- & -- & -- \\
\\
L$\alpha$ $\lambda$1216 & 55~$\pm$~20  & FO & 20.9 & -- & 28.7 & 9.62 \\
\nv\/ $\lambda$1240 & -- & -- & 1.42 & -- & 0.43 & 0.60 \\
\siiv\/ $\lambda$1397 & -- & -- & $<$0.58 & -- & 0.09 & 0.03 \\
\oiv\/] $\lambda$1402 & -- & -- & $<$0.84 & -- & 0.22 & 0.45 \\
\civ $\lambda$1549 &  12~$\pm$~8 & FO & 4.75 & -- & 3.70 & 2.47 \\

\heii\/ $\lambda$1640 & -- & -- & 1.75 & -- & 2.18 & 2.82 \\
\ciii\/ $\lambda$1909 & 5.5~$\pm$~3.7 & FO & 1.60 & -- & 1.25 & 3.63 \\
\cii\/ $\lambda$2326 & -- & -- & 0.25 & -- & 0.39 & 0.97 \\
$[$\neiv\/] $\lambda$2424 & -- & -- & 0.32 & -- & 0.19 & 0.31 \\
$[$\oii\/] $\lambda$2471 & -- & -- & 0.11 & -- & 0.16 & 0.17 \\

\mgii\/ $\lambda$2798 & 1.8~$\pm$~1.5 & FO & 0.43 & -- & 1.42 & 0.23 \\
\heii\/ $\lambda$3205 & -- & -- & 0.13 & -- & 0.12 & 0.15 \\
$[$\nev\/] $\lambda$3426 &  1.2~$\pm$~0.2 & FO & 0.36 & -- & 0.62 & 0.67 \\
$[$\oii\/] $\lambda$3727 &  3.2~$\pm$~1.7  & FO & 2.27 & -- & 2.32 & 3.20 \\
$[$\neiii\/] $\lambda$3869 &  1.4~$\pm$~0.4  &  FO & 0.81 & -- & 1.13  & 1.00 \\

$[$\suii\/] $\lambda$4074 &  0.3~$\pm$~0.2 & FO  & 0.12 & -- & 0.26  & 0.28  \\
H$\delta$ $\lambda$4102 & -- & -- & 0.18 & -- & 0.27 & 0.27\\
H$\gamma$ $\lambda$4340 & -- & -- & 0.48 & -- & 0.48 & 0.47

\\[0.01cm]
\hline
\end{tabular}
\end{center}
\end{table}

\clearpage
\begin{table}[h]
\begin{center}
\begin{tabular}{ccccccc}
\multicolumn{7}{c}{\sc TABLE~1 Continued} 
\\[0.1cm]
\multicolumn{7}{c}{\sc Observed versus integrated spectra}
\\[0.2cm]
\hline
\hline
%
\multicolumn{1}{c}{Line}
&\multicolumn{1}{c}{Mean Sy 2$^a$}
&\multicolumn{1}{c}{Reference$^b$}
&\multicolumn{1}{c}{NGC~3393$^c$}
&\multicolumn{1}{c}{Circinus$^d$}
&\multicolumn{1}{c}{Solar$^e$}
&\multicolumn{1}{c}{dusty$^f$}\\
\multicolumn{1}{c}{(1)} & \multicolumn{1}{c}{(2)} & 
\multicolumn{1}{c}{(3)} & \multicolumn{1}{c}{(4)} &
\multicolumn{1}{c}{(5)} & \multicolumn{1}{c}{(6)} & 
\multicolumn{1}{c}{(7)} 
\\[0.05cm]
\hline
$[$\oiii\/] $\lambda$4363 &  0.2~$\pm$~0.1  & FO & 0.09 & -- & 0.21 & 0.22  \\
\heii\/ $\lambda$4686 &  0.3~$\pm$~0.1  &  FO & 0.25 & 0.30 & 0.30 & 0.37  \\

H$\beta$ $\lambda$4861 & 1.00 & FO & 1.00 & 1.00 & 1.00 & 1.00 \\
$[$\oiii\/] $\lambda$5007 & 10.8~$\pm$~3.0 & FO  & 10.2 & 11.8 & 11.7  & 10.8 \\
$[$\ni\/] $\lambda$5199 &  0.15~$\pm$~0.09 & FO & 0.15 & 0.35 & 0.07  & 0.28  \\
$[$\fexiv\/] $\lambda$5303 &  -- & -- & -- & $<$0.10 & 0.033  & 0.004  \\
\hei\/ $\lambda$5876 &  0.13~$\pm$~0.06 & FO  & $<$0.11 & -- &  0.13  & 0.11  \\

$[$\fevii\/] $\lambda$6087 &  0.10~$\pm$~0.05 & FO & -- & 0.055 & 0.049  & 0.009  \\
$[$\oi\/] $\lambda$6300 &  0.57~$\pm$~0.20 & FO  & 0.35 & 0.36 &  0.56  & 0.84  \\
$[$\fex\/] $\lambda$6374 &  0.04~$\pm$~0.04 & FO & -- & 0.010 & 0.015  & 0.002  \\
H$\alpha$ $\lambda$6563 & 3.1~$\pm$~0.1  & FO & 3.00 & 2.73 & 2.96 & 2.91 \\
$[$\nii\/] $\lambda$6584 &  2.9~$\pm$~1.0  & FO & 3.64 & 3.19 & 1.18  & 1.69 \\

$[$\suii\/] $\lambda$6720 &  1.5~$\pm$~0.5  & FO & -- & 1.64 & 1.20  & 1.87 \\
$[$\ariii] $\lambda$7135 &  0.24~$\pm$~0.07 & FO  & -- & 0.16 & 0.22  & 0.27  \\
$[$\caii\/] $\lambda$7291 &  0.04~$\pm$~0.04 & VB & -- & -- & 0.16  & 0.002 \\
$[$\oii\/] $\lambda$7325 &  0.10 -- 0.46$^h$ & DS & -- & 0.073 & 0.20  & 0.21 \\
$[$\fexi\/] $\lambda$7892 &  -- & -- & -- & 0.066 & 0.068  & 0.008  \\
$[$\suiii\/] $\lambda$$\lambda$9069,9532 &  0.5 -- 4.5$^h$ & DS & -- & 0.94 &  1.40  & 1.61 \\
$[$\suviii\/] $\lambda$9915 &  -- & -- & -- & 0.044 & 0.024  & 0.022  \\
$[$\suix\/] 1.252~$\mu$m & -- & --  & -- & 0.042 & 0.11  & 0.086 \\
$[$\feii\/] 1.257~$\mu$m &  0.12~$\pm$~0.01 & T  & -- & 0.09 & 1.16  & 0.23 \\
$[$\feii\/] 1.279~$\mu$m &  0.03~$\pm$~0.01 & T  & -- & -- & 0.05  & 0.01  \\
P$\beta$ 1.2818~$\mu$m & 0.16~$\pm$~0.01$^i$ & T & -- & 0.15 & 0.16 &  0.15

\\[0.01cm]
\hline
\end{tabular}
\end{center}
\end{table}

\clearpage
\begin{table}[h]
\begin{center}
\begin{tabular}{ccccccc}
\multicolumn{7}{c}{\sc TABLE~1 Continued} 
\\[0.1cm]
\multicolumn{7}{c}{\sc Observed versus integrated spectra}
\\[0.2cm]
\hline
\hline
%
\multicolumn{1}{c}{Line}
&\multicolumn{1}{c}{Mean Sy 2$^a$}
&\multicolumn{1}{c}{Reference$^b$}
&\multicolumn{1}{c}{NGC~3393$^c$}
&\multicolumn{1}{c}{Circinus$^d$}
&\multicolumn{1}{c}{Solar$^e$}
&\multicolumn{1}{c}{dusty$^f$}\\
\multicolumn{1}{c}{(1)} & \multicolumn{1}{c}{(2)} & 
\multicolumn{1}{c}{(3)} & \multicolumn{1}{c}{(4)} &
\multicolumn{1}{c}{(5)} & \multicolumn{1}{c}{(6)} & 
\multicolumn{1}{c}{(7)} 
\\[0.05cm]
\hline

$[$\feii\/] 1.534~$\mu$m &  $<$0.02~$\pm$~0.01 & T  & -- & -- & 0.10  & 0.01  \\
$[$\feii\/] 1.644~$\mu$m &  0.09~$\pm$~0.01 & T  & -- & 0.07 & 1.12  & 0.23  \\
$[$\feii\/] 1.677~$\mu$m &  0.02~$\pm$~0.01 & T  & -- & -- & 0.07  & 0.01 \\

$[$\sivi\/] 1.96~$\mu$m &  -- & --  & -- & 0.088 & 0.088  & 0.015 \\
Br$\gamma$   2.166~$\mu$m &  -- & --  & -- & 0.027$^i$ & 0.029  & 0.026 \\
$[$\caviii\/] 2.32~$\mu$m &  -- & --  & -- & 0.065 & 0.15  & 0.002 \\
$[$\sivii\/] 2.48~$\mu$m &  -- & --  & -- & 0.15 & 0.11  & 0.020 \\
$[$\siix\/] 3.94~$\mu$m &  -- & --  & -- & 0.19 & 0.17  & 0.027 \\
Br$\alpha$   4.052~$\mu$m &  -- & --  & -- & 0.079$^i$ & 0.080  & 0.073 \\
$[$\feii\/] 25.98~$\mu$m &  -- & -- & -- & -- & 0.64  & 0.13     

\\[0.01cm]
\hline
\end{tabular}
\end{center}
\rm
\footnotesize
\leftskip=2.em
$^a$The range of observed values taken from the reference in column 3.\\
\leftskip=2.em
$^b$reference of the observed values: (N) is Netzer 1990, (BFP) is
Binette, Fosbury, \& Parker (1993), (FO) is Ferland \& Osterbrock
(1986), (VB) is from Villar-Mart\'{\i}n \& Binette (1997), (DS) denotes
a range of values from Dopita \& Sutherland (1995), and (T) is from an
observation of NGC~4151 by Thompson (1995).\\
\leftskip=2.em
$^c$Observations of NGC~3393 from Cooke et~al.\ (1997).\\
\leftskip=2.em
$^d$Observations of the Circinus galaxy from Oliva et~al.\ (1994).\\
\leftskip=2.em
$^e$Predictions from the integrated spectrum assuming solar abundances 
indicated by the square in Fig.~3.\\
\leftskip=2.em
$^f$Predictions from the integrated spectrum assuming \hii\/ region abundances indicated by the star in Fig.~3.\\
\leftskip=2.em
$^g$Equivalent width of H$\beta$ in \AA\/.\\
\leftskip=2.em
$^h$These observations are given as a range.\\
\leftskip=2.em
$^i$Assuming Case B conditions.

\end{table}
%
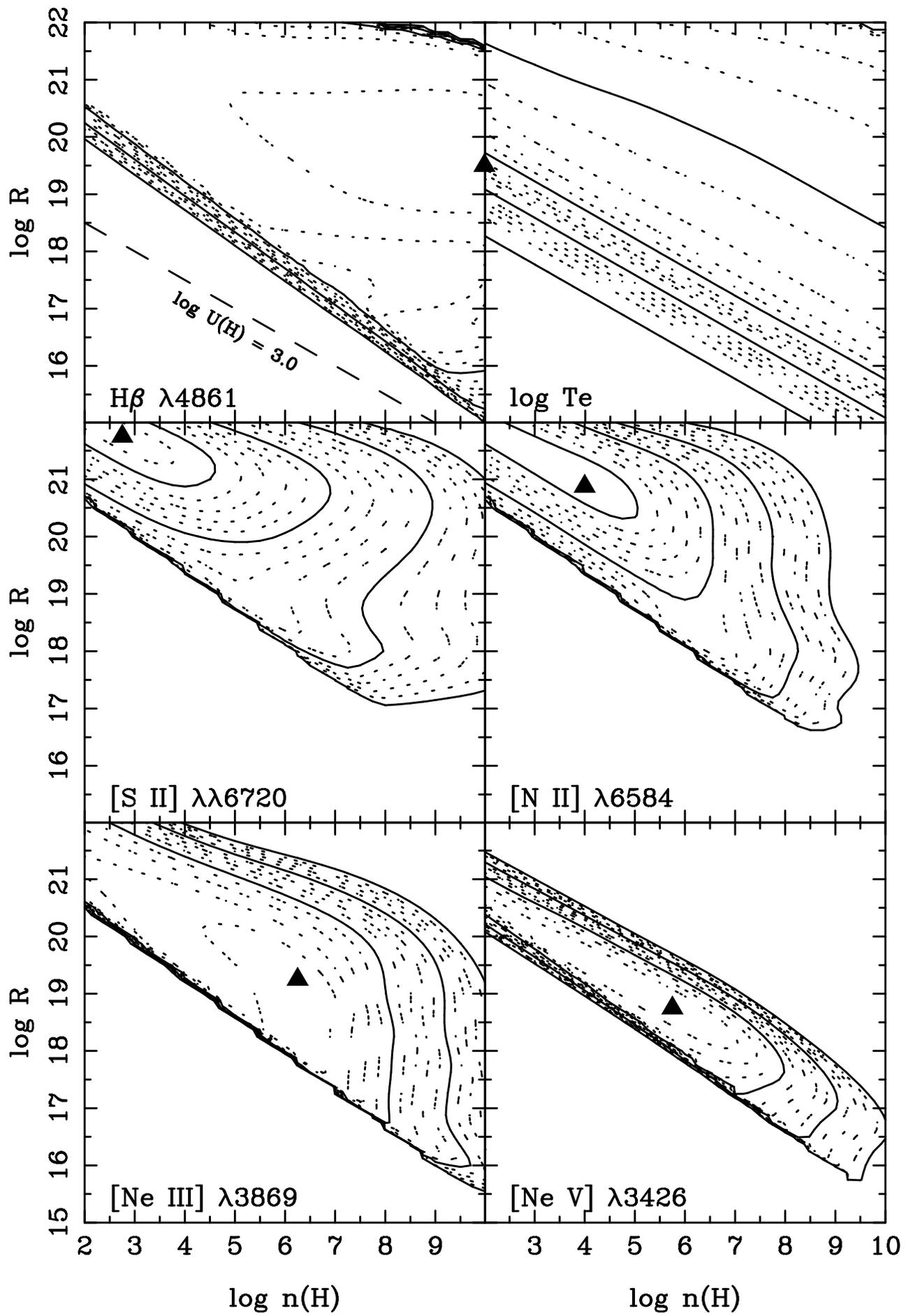
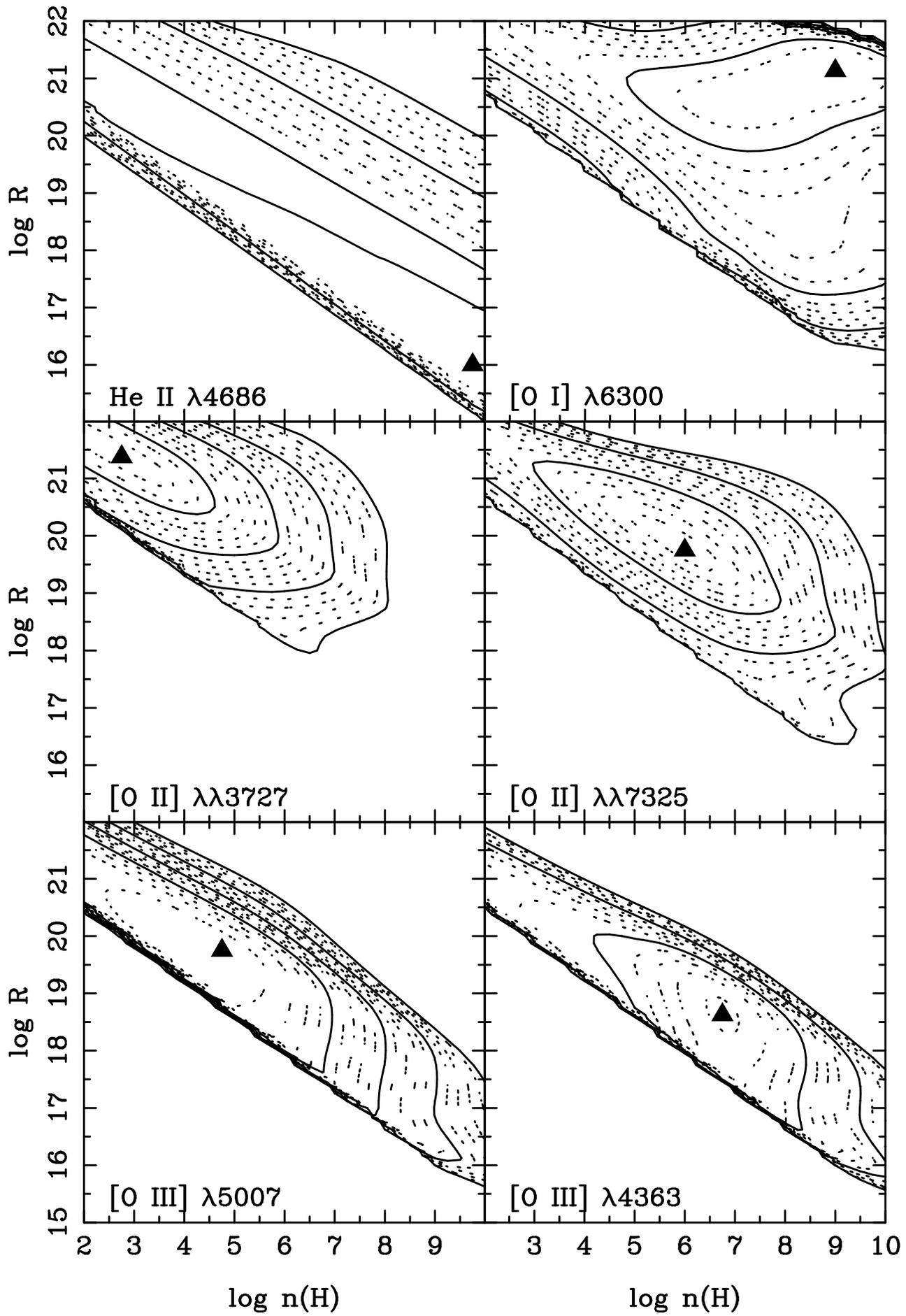
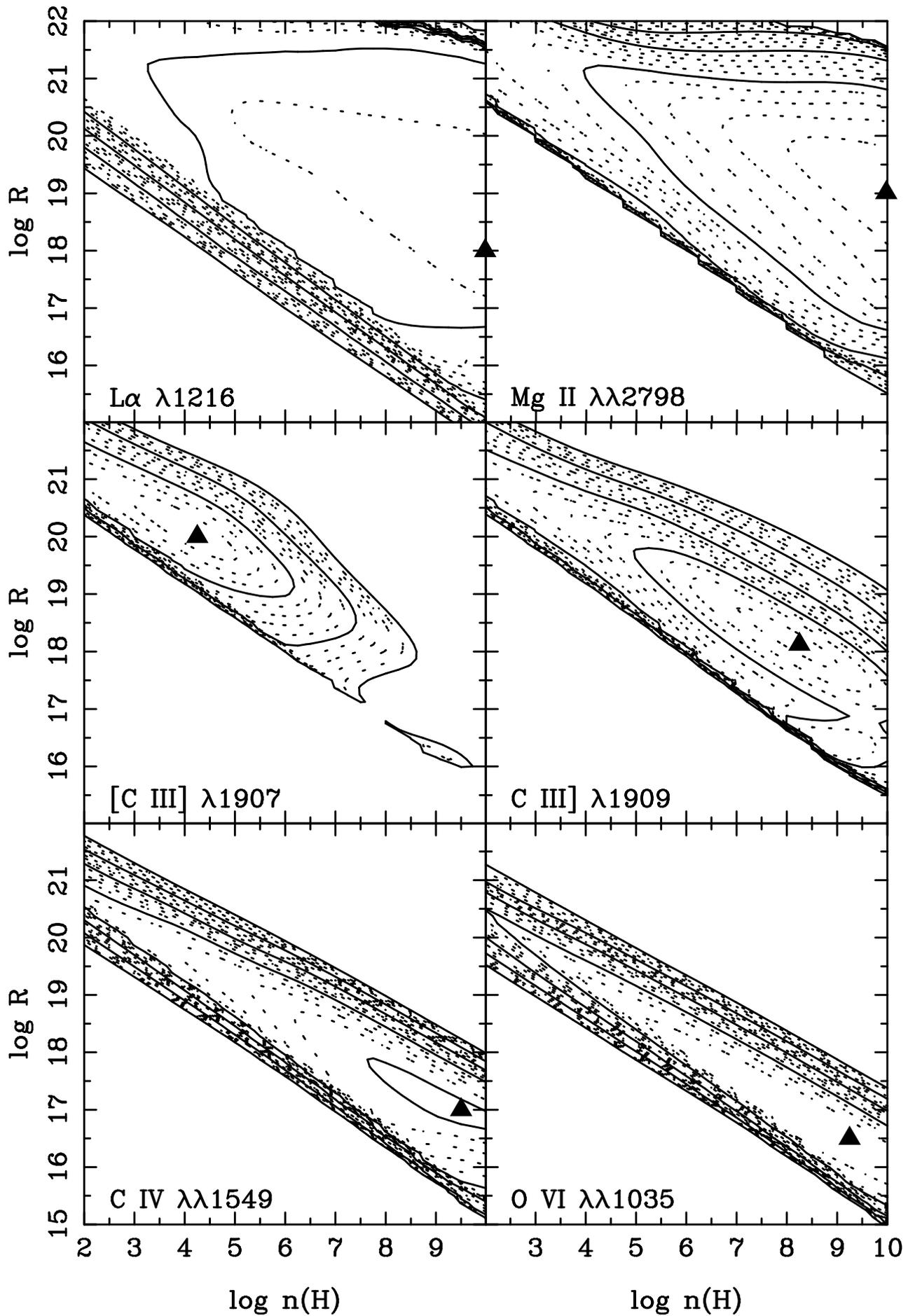
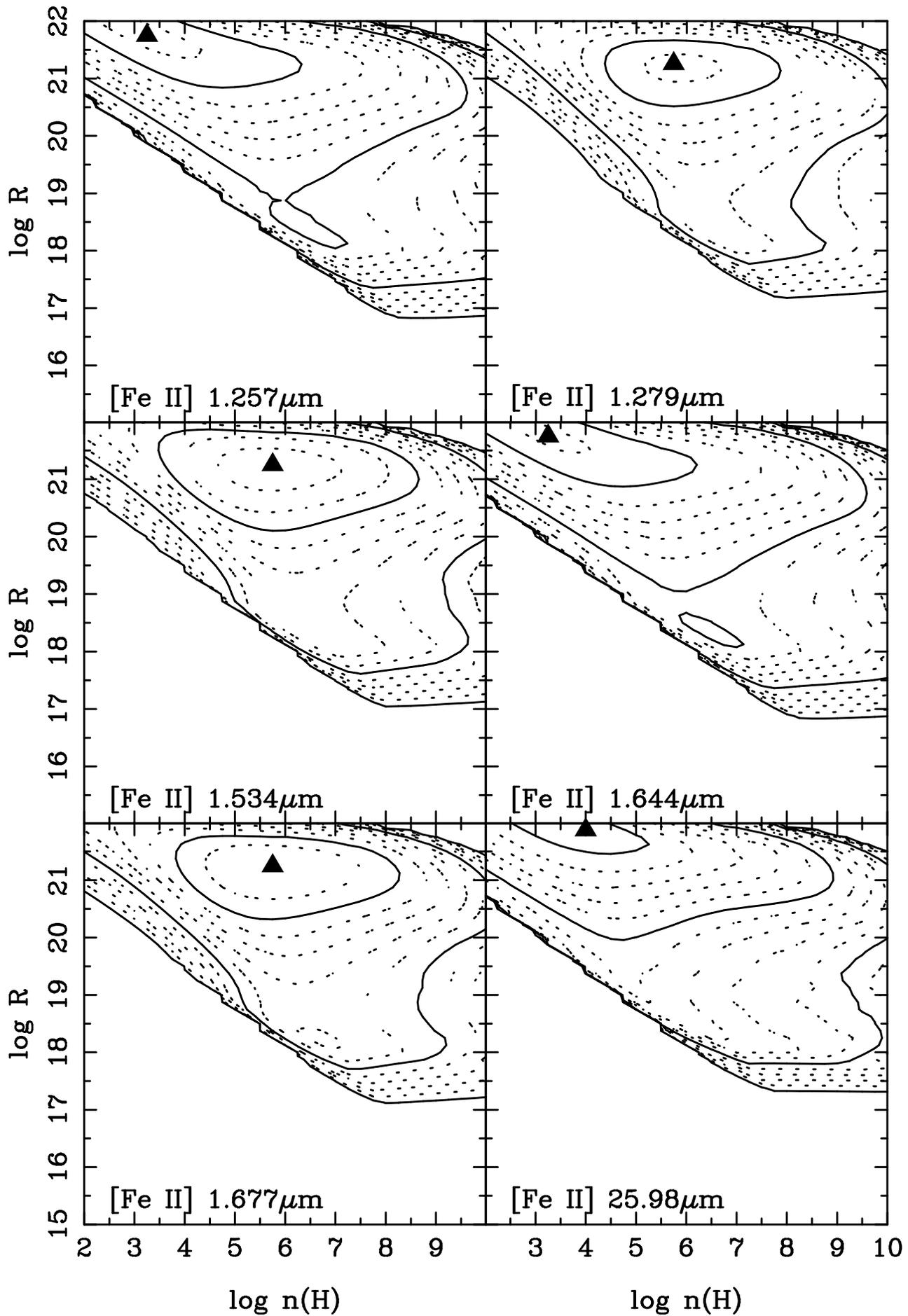
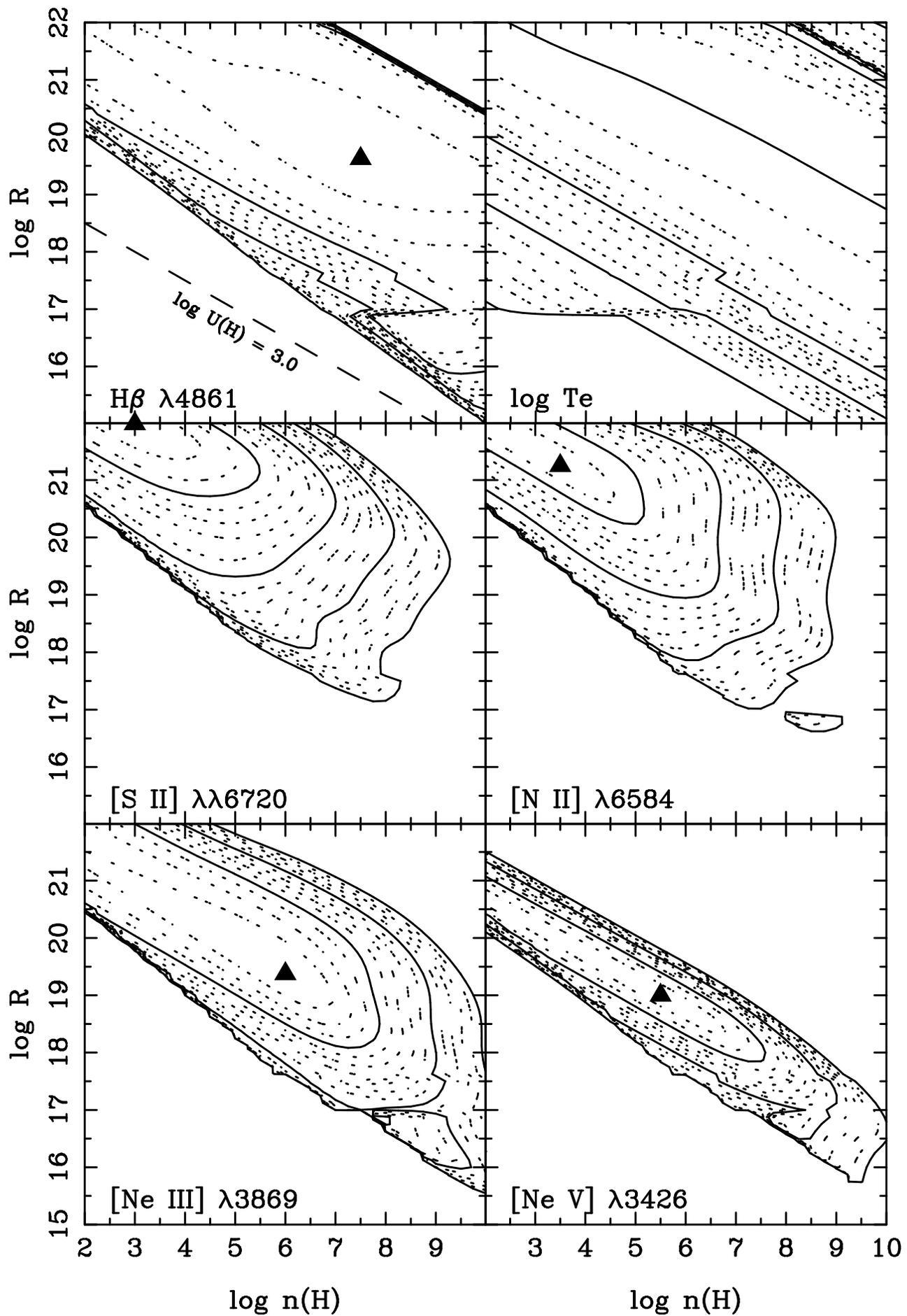
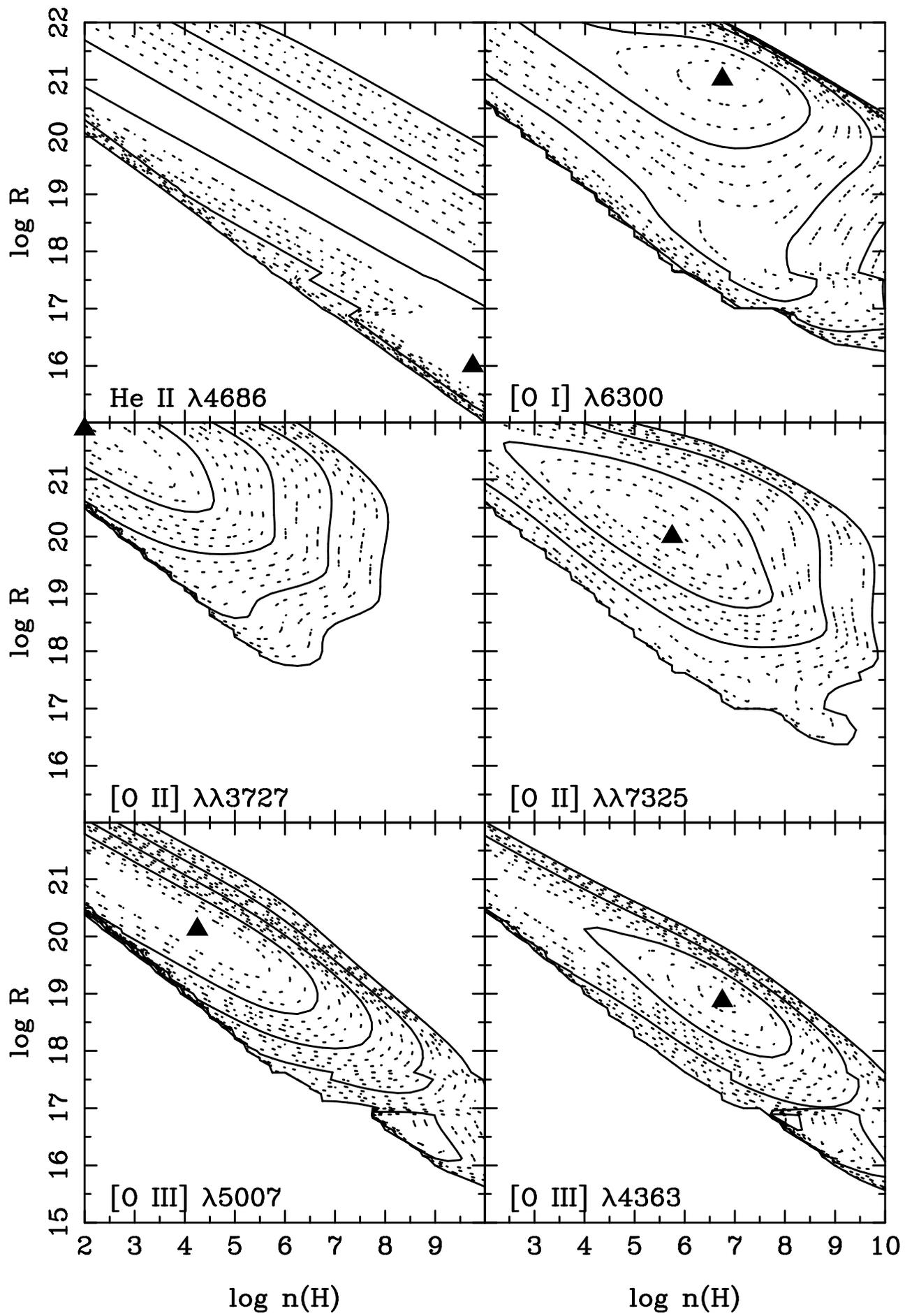
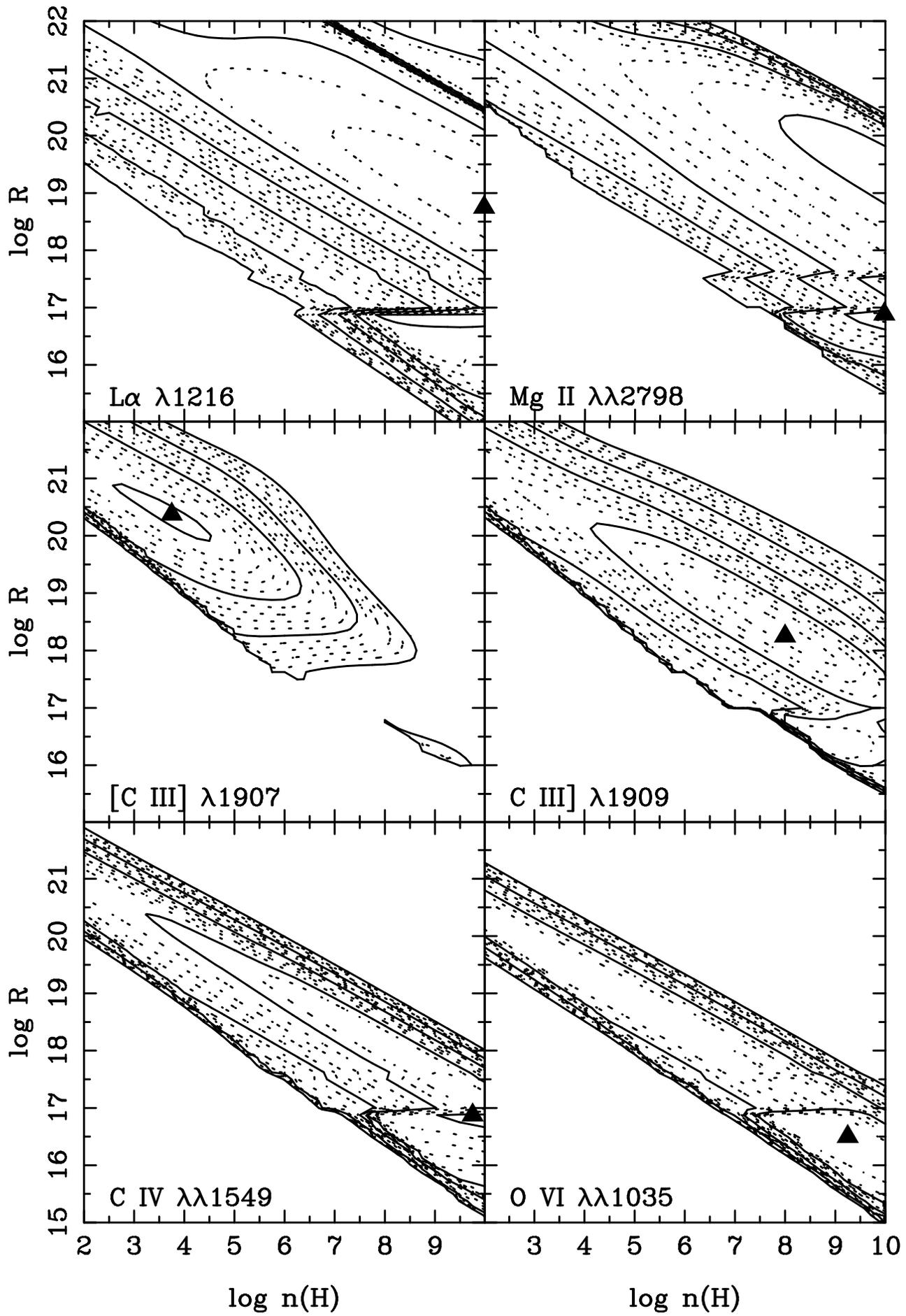
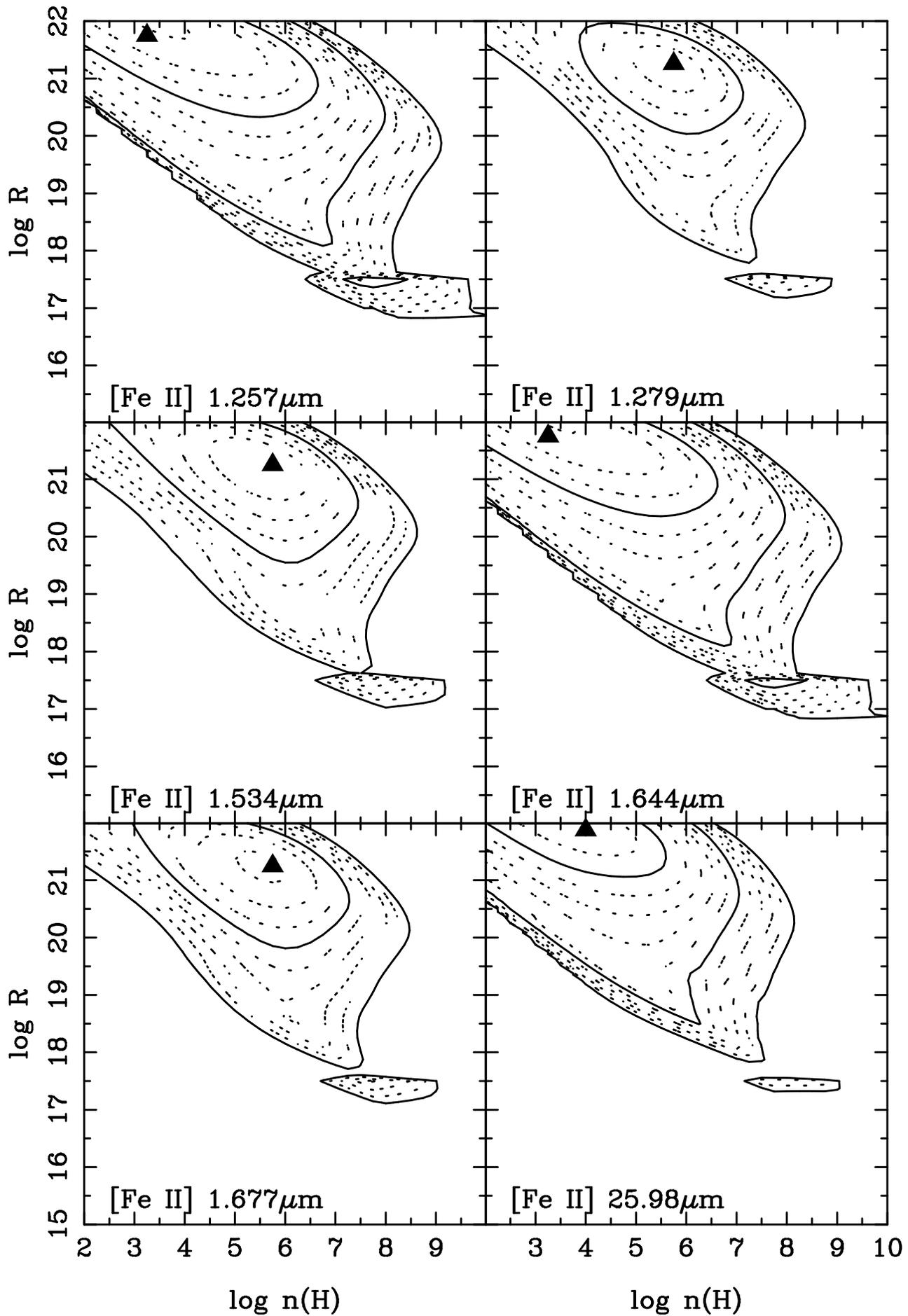
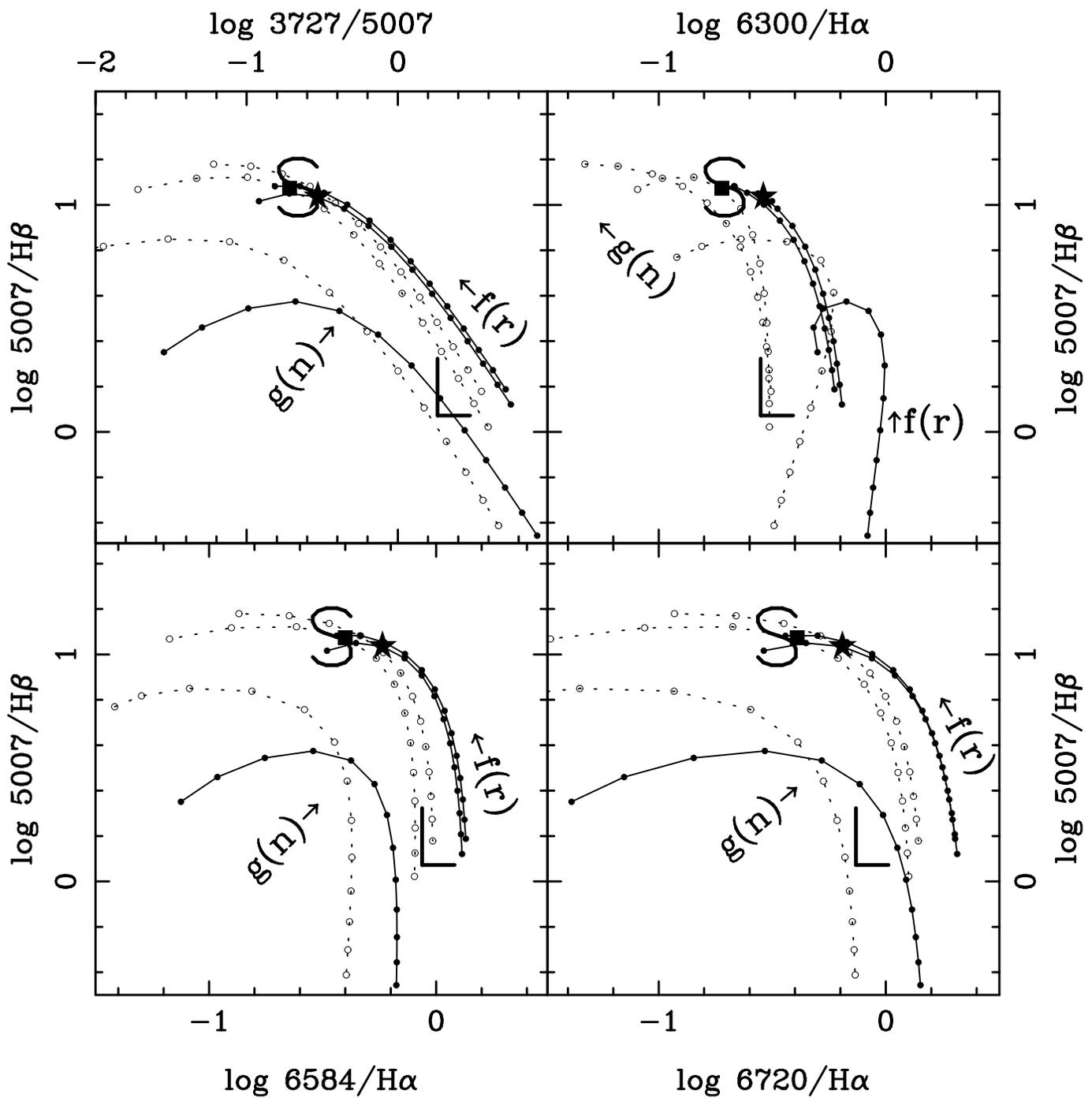
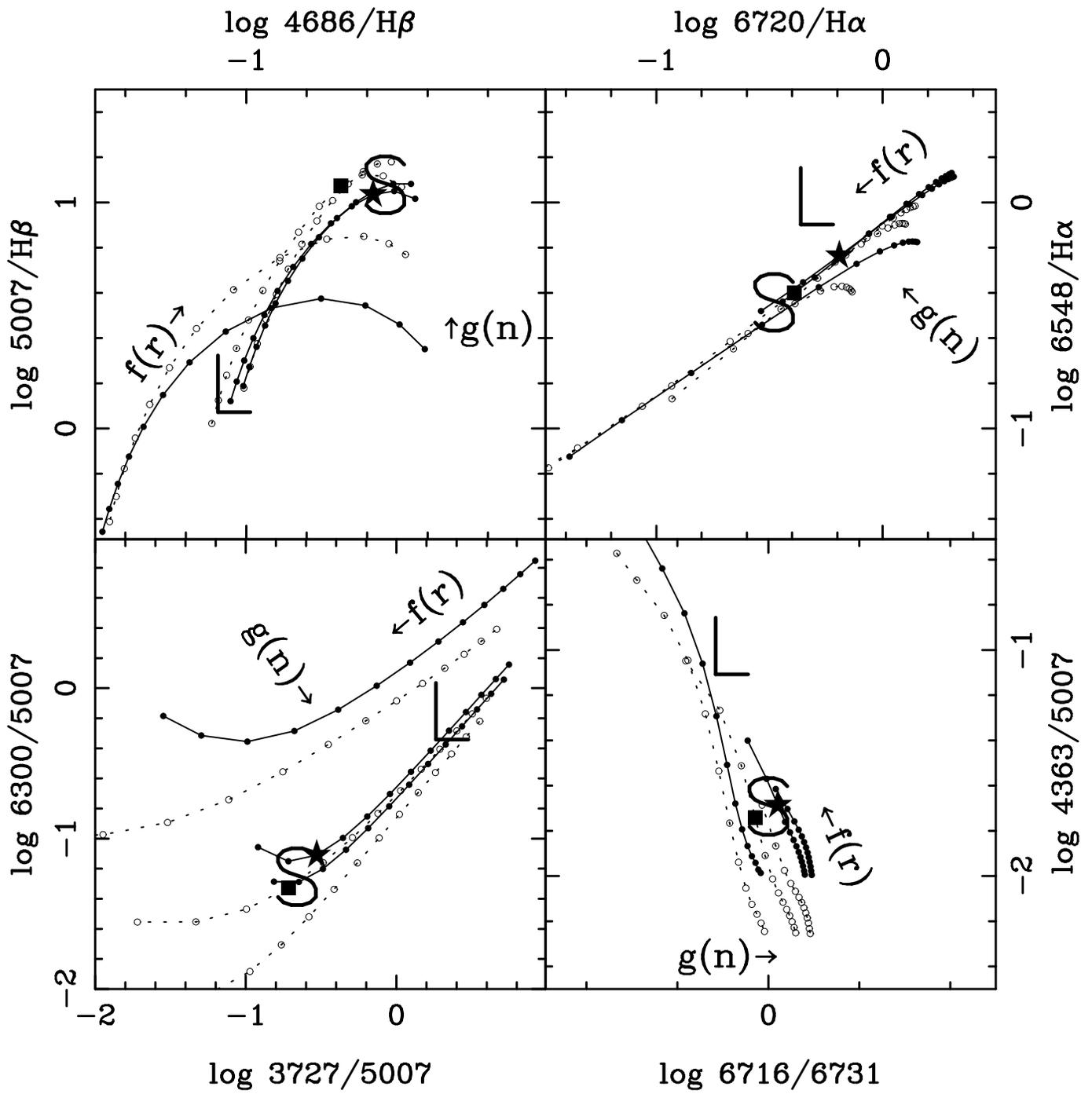
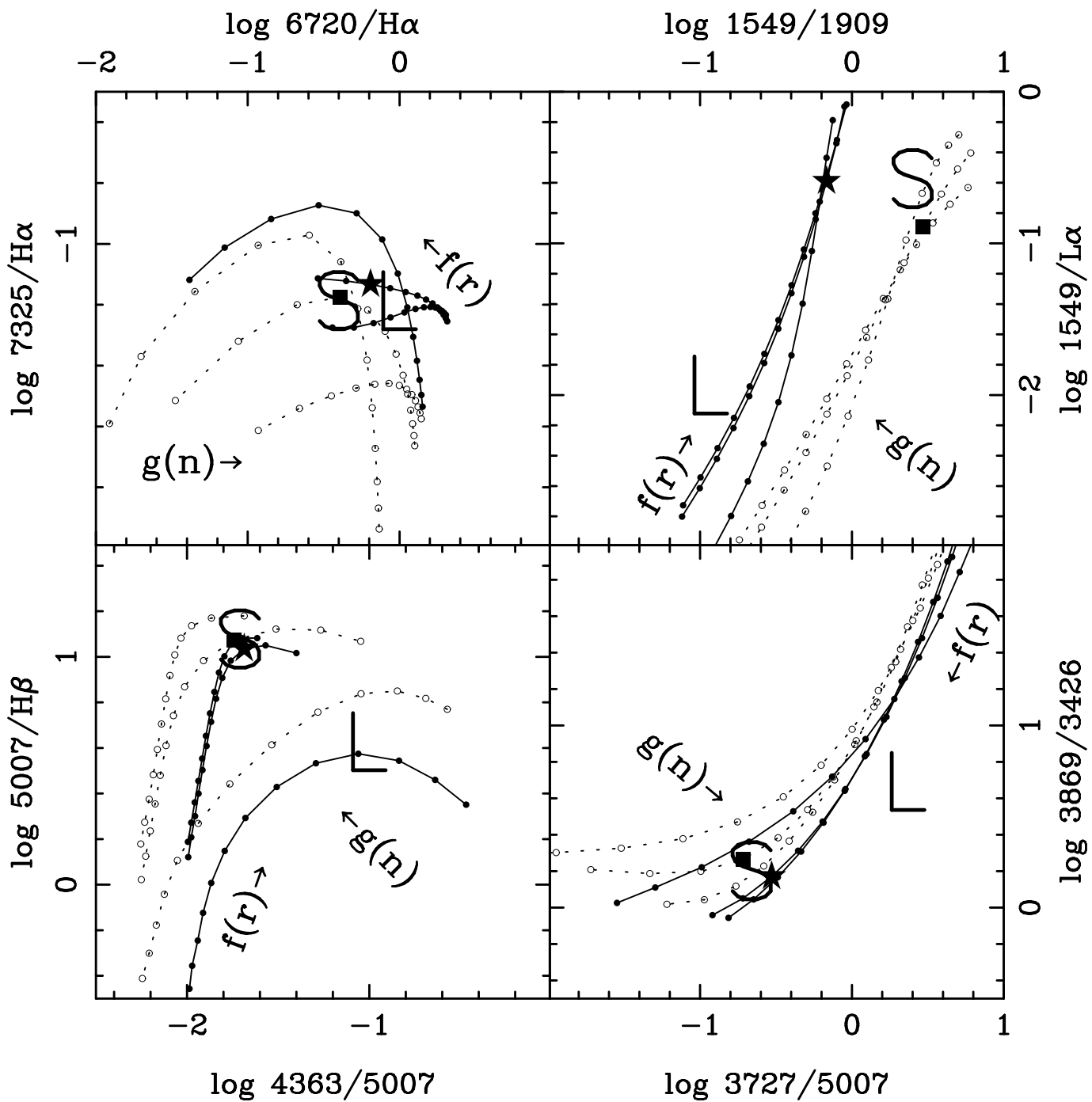
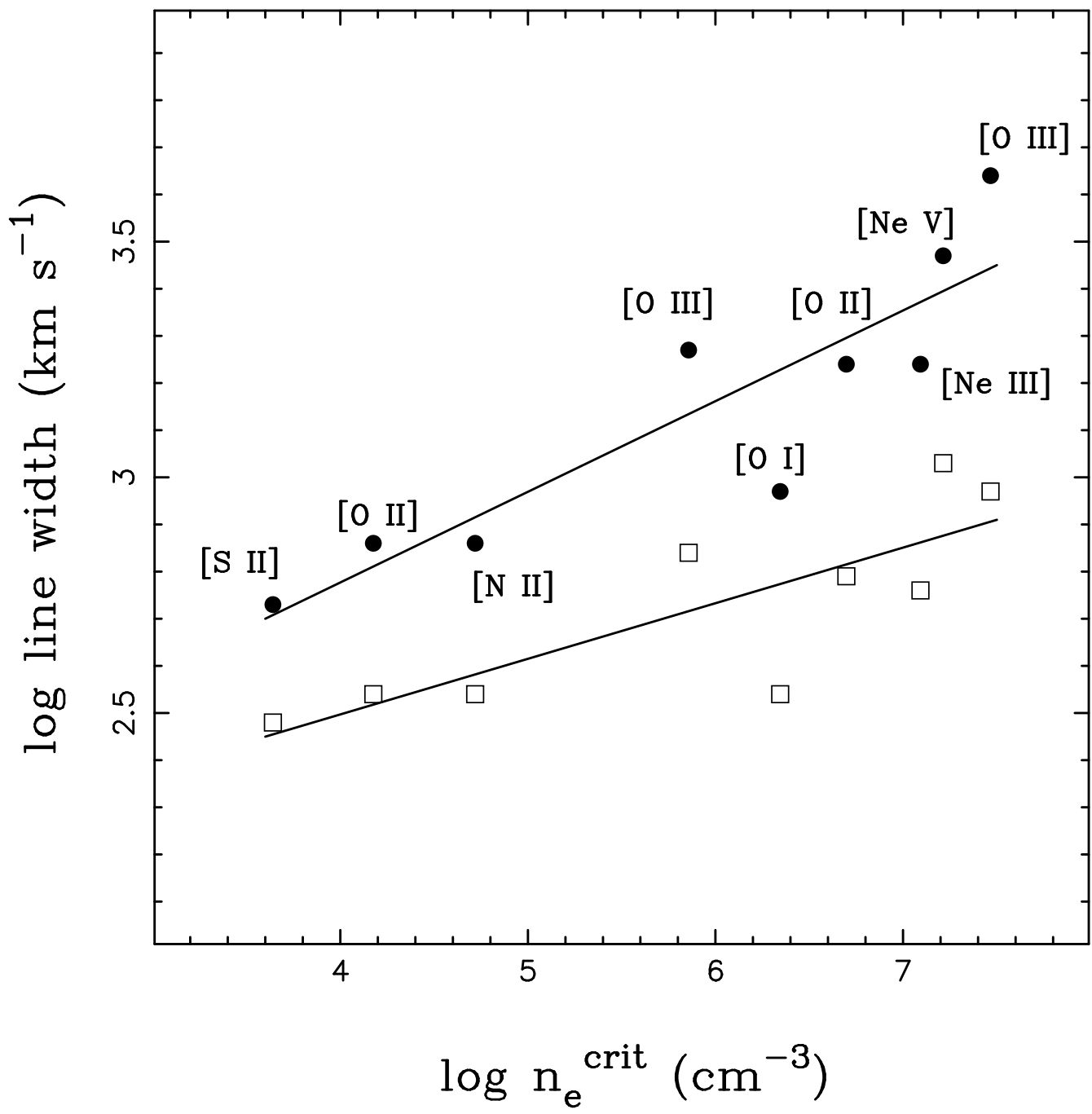
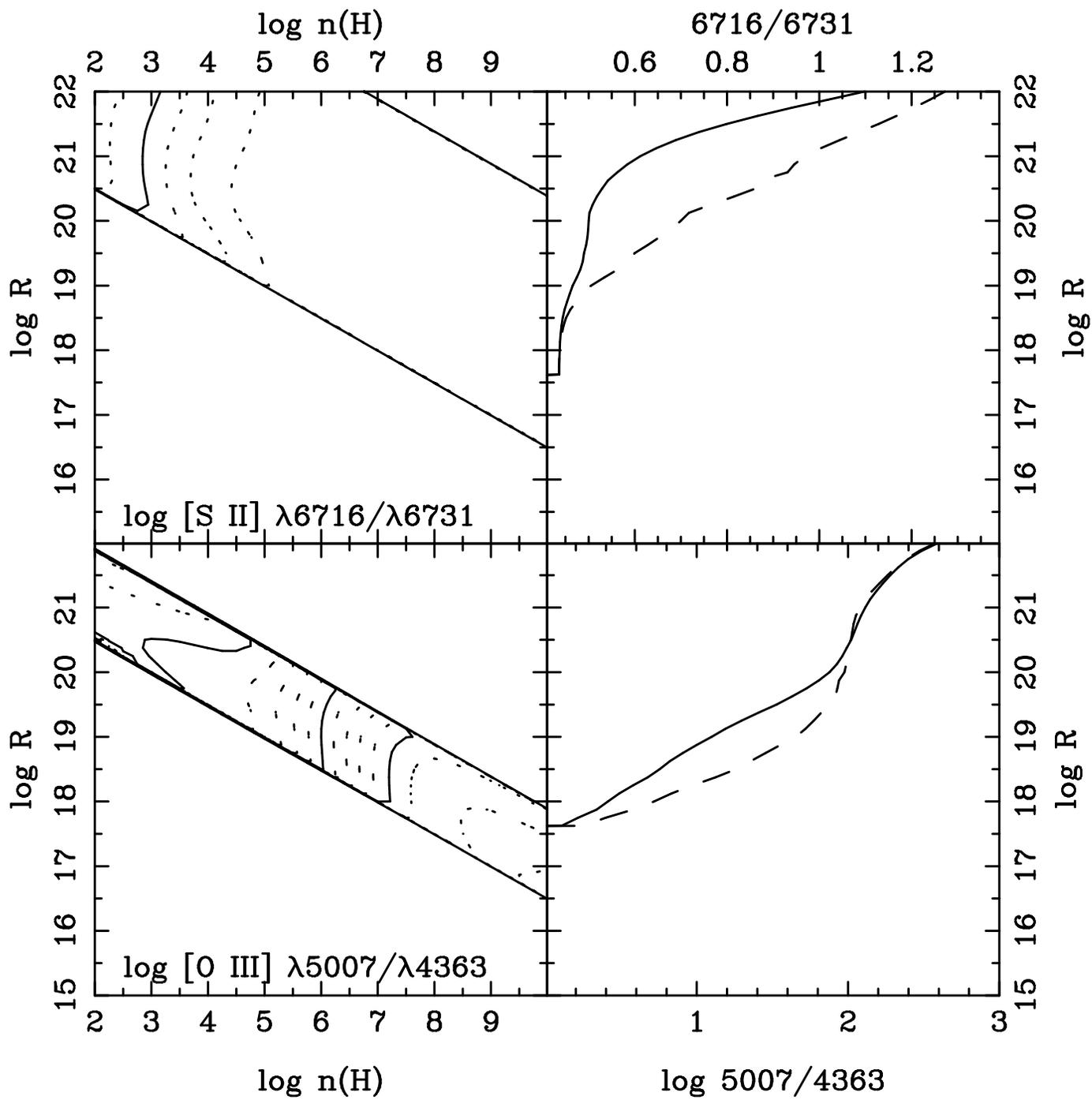
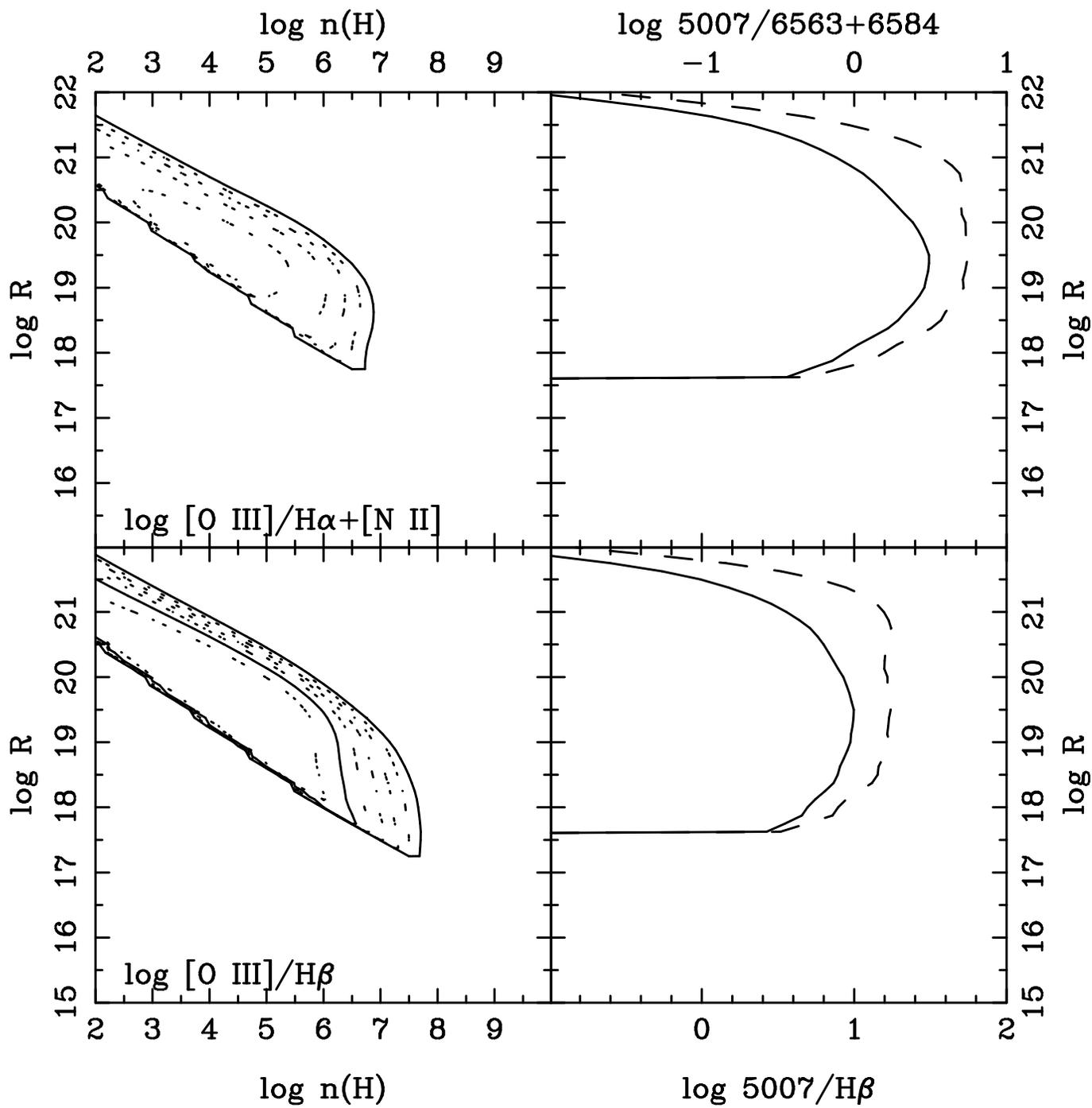
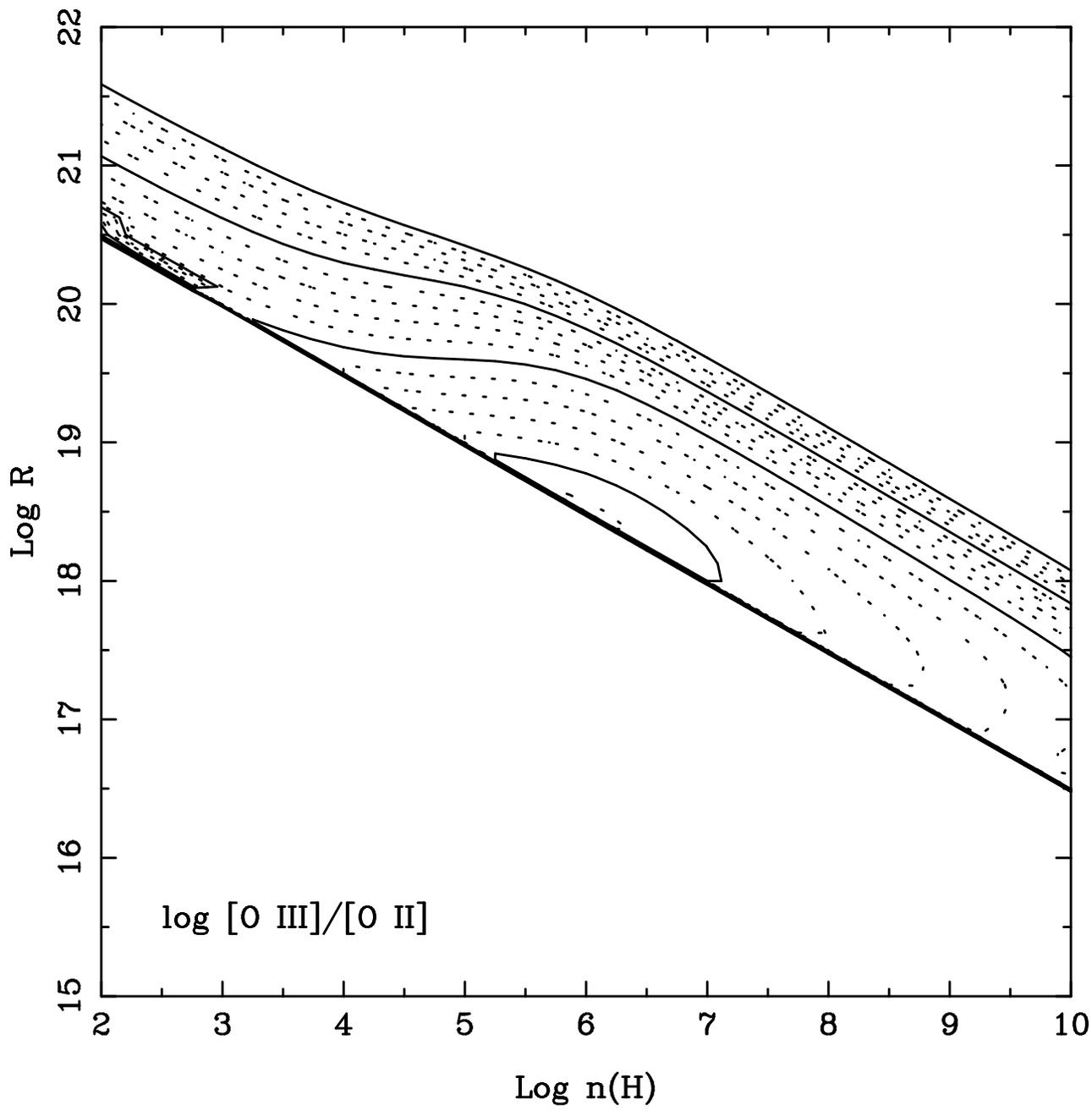
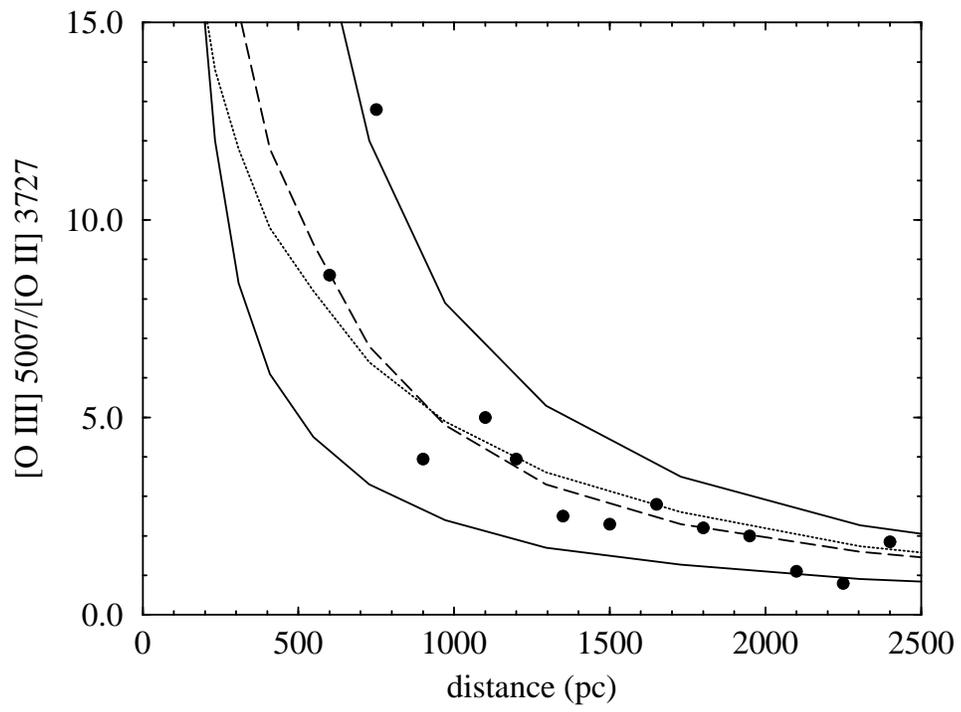
\end{document}